\documentclass[10pt,journal]{IEEEtran}
\pdfoutput=1

\usepackage[pdftex]{graphicx}

\usepackage[cmex10]{amsmath}
\usepackage{algorithmic}
\def\sval{\sigma}
\def\svalmax{\sval_\text{max}}
\def\svalmin{\sval_\text{min}}
\def\nv{N_\text{views}}
\def\nb{N_\text{bins}}
\DeclareMathOperator*{\argmin}{argmin}

\def\ltwotv{$\ell_2$-TV}
\def\ltwomag{$\ell_2$-magnitude}
\def\ltworough{$\ell_2$-roughness}
\newcommand{\TV}[1]{\| #1 \|_\mathrm{TV}}

\begin{document}
%
\title{
Quantifying admissible undersampling \\ for sparsity-exploiting iterative image reconstruction in X-ray CT
}
%
%
%


\author{Jakob~H.~J{\o}rgensen, \thanks{Jakob H. J{\o}rgensen is with the Department of Informa\-tics and Mathematical Modeling, Technical University of Denmark, Richard Petersens Plads, Building 321, 2800 Kgs. Lyngby, Denmark.
Corresponding author: Jakob H. J{\o}rgensen, e-mail:
jakj@imm.dtu.dk.}%
        Emil Y. Sidky,  \thanks{Emil Y. Sidky and Xiaochuan Pan are with the Department of Radiology, University of Chicago, 5841 S. Maryland Ave., Chicago, IL 60637, USA, e-mail: \{sidky,xpan\}@uchicago.edu.}%
        and Xiaochuan Pan}

\maketitle

\begin{abstract}
Iterative image reconstruction (IIR) with sparsity-exploiting
methods, such as total variation (TV) minimization, investigated in compressive sensing (CS) claim potentially
large reductions in sampling requirements. Quantifying this
claim for computed tomography (CT) is non-trivial, because
both full sampling in the discrete-to-discrete imaging model and the reduction in sampling admitted by sparsity-exploiting
methods are ill-defined. 
The present article proposes definitions of full sampling by introducing four sufficient-sampling conditions (SSCs).
The SSCs are based on the condition number of the system matrix of a linear
imaging model and address invertibility and stability. 
In the example application of breast CT, the SSCs are used as reference points of full sampling for quantifying the undersampling admitted by reconstruction through TV-minimization.
In numerical simulations, factors affecting admissible undersampling are studied. Differences between few-view and few-detector bin reconstruction as well as 
a relation between object sparsity and admitted undersampling are quantified.
\end{abstract}


%
\IEEEpeerreviewmaketitle

%
%
%
%
%
%


\section{Introduction}

\IEEEPARstart{R}{ecently}, iterative image reconstruction (IIR) 
algorithms have been developed for X-ray tomography
\cite{SidkyTV:06,song2007sparseness,chen2008prior,sidky2008image,sidky2009enhanced,PanIP:09,
Choi:10,Bian:10,ritschl2011improved,han2011algorithm,Defrise:11}
based on the ideas discussed in the field of compressive sensing (CS)
\cite{Donoho2006,candes2006robust,candes2008introduction,candes2006stable}.
These algorithms promise accurate reconstruction from less data than is required by standard image
reconstruction methods.
This is made possible by exploiting sparsity, i.e.,
few non-zeroes in the image or of some transform applied to the image.
One can argue about whether these algorithms are truly novel or not: edge-preserving
regularization and reconstruction based on the total variation (TV)
semi-norm \cite{rudin1992nonlinear,delaney1998globally,persson2001total,elbakri2002statistical}
have a clear link to sparsity in the object gradient and have been considered before the advent of CS,
and algorithms specifically
for object sparsity have been developed for blood vessel imaging with contrast agents
\cite{li2002accurate}.
Nevertheless, the interest in CS has
broadened the perspective on applying optimization-based
methods for IIR algorithm development for computed tomography (CT),
and it has motivated development of
efficient algorithms involving variants of the $\ell_1$-norm
\cite{Defrise:11,Beck2009a,Jensen2012,ChambollePock:2011,Ramani2012,Rashed2012,Sidky2012}.

What is seldom discussed, however, is that the theoretical results from CS do not
extend to the CT setting. CS only provides theoretical guarantees of accurate undersampled
recovery for certain classes of random measurement matrices \cite{candes2008introduction},
not deterministic matrices such as CT system matrices.
While the mentioned references demonstrate empirically
that CS-inspired methods do indeed allow for undersampled CT reconstruction,
there is a fundamental lack of understanding of why, and the conditions under which, this is the case.
One problem in uncritically applying sparsity-exploiting methods to CT is that there is no
quantitative notion of full sampling.

Most IIR, including sparsity-exploiting, methods employ a discrete-to-discrete (DD) imaging 
model\footnote{See \cite[Chapter 15]{Barrett:FIS} for an overview of different imaging models.}
which requires that the
object function be represented by a finite-sized expansion set and sampling
specified over a finite set of transmission rays. 
This contrasts with most analysis of CT sampling, which is performed on a
continuous-to-discrete (CD) model. 
For analyzing analytic
algorithms such as filtered back-projection (FBP), 
a continuous-to-continuous (CC) model
such as
the X-ray or Radon transform is chosen, and discretization of the data space
is considered, yielding results for the corresponding CD model.
Analysis of the CD model is performed independent of object expansion.
If the expansion set for the DD model is chosen to be point-like, e.g. pixels/voxels,
there may be similarity between CD and DD models justifying some
crossover of intuition on sampling, but in 
general 
sufficient-sampling conditions can be
different for the two models. That a more fine-grained notion of sufficient sampling is
needed for the DD model can be seen by considering the representation of the object function on
a 128$\times$128 versus a 1024$\times$1024 pixel array. Clearly, the latter case 
requires more samples than the former, but sampling conditions derived
from the CD model cannot make this distinction. Sufficient sampling for the DD model becomes even
less intuitive for non-point-like expansion sets such as natural pixels, wavelets, or harmonic
expansions. Yet, to quantify the the level of \emph{under}sampling admitted by a sparsity-exploiting
IIR method, 
\emph{full} sampling needs to be defined 
for the corresponding DD model, 
and to that end we introduce several sufficient-sampling conditions (SSCs).

Specifically, in the present article, SSCs for the DD model are derived from the
condition number of the corresponding system matrix. Multiple SSCs are defined to
characterize both invertibility and stability of the system matrix.
To perform the analysis, a class of 
system matrices 
is defined so that the system matrix depends on few parameters. The class is chosen so that
it has wide enough applicability to cover thoroughly a configuration/expansion combination of interest, but
not so wide as to make the analysis impractical. For the present study, we select a 
system matrix class
for a 2D circular fan-beam geometry using a square-pixel array.
The SSCs are chosen so that they provide a useful characterization of any 
system matrix class, but
the particular values associated with the SSCs in this work apply only to the narrow
system matrix class defined. While the article presents a strategy for 
defining full sampling, 
the analysis must be redone with any alteration to the 
system matrix class.

After deriving the SSCs for the particular circular fan-beam CT 
system matrix class, we apply
sparsity-exploiting IIR in the form of constrained TV-minimization. We consider the specific application of CT to breast imaging and use a realistic and challenging discrete phantom.
We use the SSCs as reference points of full sampling for quantifying the undersampling admitted by each of the conducted reconstructions. Specifically, we demonstrate significant differences in undersampling admitted for reconstruction from few views compared to few bins. We study how variations to the reconstruction optimization problem, to the image quality metric, to the discretization method for the system matrix, and to the sparsity of the phantom image affect the results.

In Sec. \ref{sec:model} we describe the CT imaging model and
present the particular 
system matrix class we employ for circular fan-beam CT reconstruction. In 
Sec. \ref{sec:CS} we give a background on sparsity-exploiting methods. 
In Sec. \ref{sec:svd} the SSCs are presented 
and their application 
is illustrated for the 2D circular fan-beam case. Finally,
Sec. \ref{sec:numerical} illustrates an example study on quantifying admissible undersampling
by constrained TV-minimization employing the discussed SSCs.


\section{A class of system matrices for the discrete-to-discrete imaging model}
\label{sec:model}

\subsection{The X-ray transform}

Explicit image reconstruction algorithms such as FBP
are based on inversion formulas 
for the CC cone-beam or X-ray transform model,
\begin{equation}
\label{conebeam}
g[\vec{s},\vec{\theta}] = \int_0^\infty dt f( \vec{s} + t \vec{\theta}),
\end{equation}
where $g$, the line integral over the object function $f$ from source location
$\vec{s}$ in the direction $\vec{\theta}$, is considered data. 
Fan-beam FBP, for example, inverts this model for the case where the source location
$\vec{s}$ varies continuously on a circular trajectory surrounding the subject, and
at each $\vec{s}$ the ray-direction $\vec{\theta}$ is varied continuously through the
object in the plane of the source trajectory. 

\subsection{The discrete-to-discrete model}

For most IIR algorithms, the CC imaging model
is discretized by expanding the object function in a finite expansion set, for example,
in pixels/voxels. Furthermore, the discrete digital sampling of the CT device is accounted
for by directly using the sampled data without interpolation.
The effect of both of these steps is to convert the imaging model to a discrete-to-discrete (DD)
formulation,
\begin{equation}
\label{xfeg}
\vec{g} = X \vec{f},
\end{equation}
where $\vec{g}$ represents a finite set of ray-integration samples, $\vec{f}$ are coefficients
of the object expansion, and $X$ is the system matrix modeling ray integration. This
DD imaging model is almost always solved implicitly, because the matrix $X$, even though sparse, is
beyond large for CT applications: $X$ is in the domain
of a giga-matrix for 2D imaging and a tera-matrix for 3D imaging.

A central point motivating the strategy of the present work is that
the DD imaging model 
has narrower scope than the CD model, because it often
derives from the CD model by
expanding the continuous image domain with a finite set of functions.
How the discretization of the CD model is done for CT
to achieve the DD imaging model is not standardized.
Many expansion elements have been used in CT studies; in addition to pixels, for example blobs \cite{matej1996practical},
wavelets \cite{delaney1995multiresolution,unser1996review}, and
natural pixels \cite{buonocore1981natural,gullberg1994reconstruction}.
Also,
the matrix elements using only the pixel expansion set can be calculated in different ways that
all tend toward the CC model in the limit of shrinking pixel size and detector-bin size.
Different modeling choices will necessarily alter $X$.
This tremendous variation in $X$ means that it is important to fully specify
$X$ for each study, and it is important to re-characterize $X$ for
any change in the model. For example, changing pixel size can have large impact on the 
null space of the system matrix in the DD model. 

In order to describe precisely and provide a delimitation of the system matrices considered in the present work, 
we introduce the notion of a \emph{system matrix class}.
Any given system matrix depends on numerous model parameters determining the scanning geometry, sampling and discrete expansion set. A system matrix class consists of the system matrices arising from fixing a number of these parameters and leaving a subset of the parameters free. The system matrix class can then be studied by varying these free parameters.

\subsection{The system matrix class used in the present study}

In CT,
projections are acquired 
from multiple
source locations which lie on a curve trajectory and the source
location $\vec{s}(\lambda)$ is specified by the scalar parameter $\lambda$.
The circular trajectory is the most common, and is what we use here,
\begin{equation}
\vec{s}(\lambda) = R_0 (\cos \lambda, \sin \lambda),
\notag
\end{equation}
where $R_0$ is the distance from the center-of-rotation to the X-ray source, and set to $R_0 = 40$ cm in the present work.
The detector bin locations are given by
\begin{equation}
\vec{b}(\lambda,u) = (R_0-D) (\cos \lambda, \sin \lambda) + u (- \sin \lambda, \cos \lambda) ,
\notag
\end{equation}
where $D$ is the source-to-detector-center distance ($D = 80$ cm in the present work), and $u$ specifies a position on the detector.
The ray direction for the detector-geometry independent data function is
\begin{equation}
\notag
\vec{\theta}(\lambda,u) = \frac{\vec{b}(\lambda,u) - \vec{s}(\lambda)}
{\|\vec{b}(\lambda,u) - \vec{s}(\lambda)\|_2}.
\end{equation}
The $2 \pi$ arc is divided into $N_\text{views}$ equally spaced angular intervals, so that the source parameters follow
\begin{equation}
\label{sampling2Dviews}
 \lambda_i=i \Delta \lambda, \text{ where }
\end{equation}
\begin{equation}
  \Delta \lambda = 2 \pi / N_\text{views}\text{ and }
i \in [0, N_\text{views} - 1].
\end{equation}
The detector is subdivided into $N_\text{bins}$,
\begin{equation}
\label{sampling2Dbins}
u_j = u_\text{min} + (j+0.5) \Delta u,
\end{equation}
where $D_L$ is the detector length ($D_L = 41.3$ cm), $u_\text{min}= -D_L/2$,
$\Delta u = D_L/N_\text{bins}$, and $j \in[0,N_\text{bins} - 1]$.
The detector length is determined by requiring it to detect all rays passing through
the largest circle inscribed within the square $N \times N$ image array for which we use the side length $20$ cm.
We restrict the unknown pixel values to lie within this circular field-of-view (FOV), 
and the number
of unknown pixel values $N_\text{pix}$ is
\begin{equation}
\label{npix}
N_\text{pix} \approx \frac{\pi}{4} N^2,
\end{equation}
where the actual value, which has to be an integer, is given with
each simulation below.

Effectively,
the dimensions of the 
projector $X$ are $M = N_\text{views} \times
N_\text{bins}$ rows (number of ray integrations) and $N_\text{pix}$ columns (number of 
variable pixels). To obtain the individual matrix elements, the line-intersection method is
employed, where $X_{m,n}$ is the intersection length of the $m$th ray with
the $n$th pixel. This description completely specifies the system matrix class for the present circular
fan-beam CT study, and the free parameters of this class are $N$, $N_\text{views}$, and $N_\text{bins}$.

\section{CT image reconstruction by exploiting gradient-magnitude sparsity}
\label{sec:CS}

Reconstruction of
objects from undersampled data within the DD imaging model corresponds to a
measurement matrix $X$ with fewer rows than columns. The infinitely many solutions are
narrowed down by selecting the sparsest one, i.e., the one that has the fewest number of
non-zeroes, either in the image itself or after some transform has been applied to it.
Mathematically, the reconstruction can be written as the solution of the constrained optimization problem
\begin{equation}
\label{L0}
\vec{f}^* = \argmin_{\vec{f}} \; \| \Psi (\vec{f}) \|_0 \text{ such that } X \vec{f} = \vec{g}.
\end{equation}
Here, $\Psi$ is a sparsifying transform, for instance a discrete wavelet transform, and
$\|\cdot\|_0$ is the $\ell_0$-``norm'' (although it is in fact not a norm), which computes its argument vector's sparsity, that is, counts the number of non-zeroes. The equality constraint restricts image
candidates to those agreeing exactly with the data.

Central results in CS derive conditions on $X$ drawn from certain random system matrix classes
such that $\vec{f}^*$ is exactly equal to the underlying unknown image that gave rise to the data $\vec{g}$. 
Two key elements are sparsity of $\Psi (\vec{f})$ and incoherence of
$X$: exact recovery depends on the size of $\vec{g}$ being larger than some
small factor of $\| \Psi (\vec{f}) \|_0$
\cite{candes2008introduction}, and
the concept of incoherence is needed to ensure that the few measurements $\vec{g}$ available give meaningful information
about the non-zero elements of $\Psi (\vec{f})$.
Other important results in CS involve the relaxation
of the non-convex $\ell_0$-``norm'' to the convex $\ell_1$-norm,
\begin{equation}
\label{L1}
\vec{f}^* = \argmin_{\vec{f}} \; \| \Psi (\vec{f}) \|_1 \text{ such that } X \vec{f} = \vec{g}.
\end{equation}
In contrast to \eqref{L0}, this convex problem is 
amenable to solution by a variety of practical algorithms, although the large scale of CT matrices still presents a challenge for
algorithm development.
Another important contribution from CS is the derivation of conditions under which the
solution to \eqref{L1} is identical to the solution to \eqref{L0}, so
that the sparsest solution can be found by solving \eqref{L1}.

For application to medical imaging, it was suggested in \cite{candes2006robust}
that a potentially useful $\Psi$ would be to have $\Psi$ compute the discrete
gradient magnitude of $\vec{f}$, i.e., for the $j$th pixel
\begin{equation}
\left[\Psi (\vec{f})\right]_j =  \|D_j\vec{f}\|_2,
\end{equation}
where $D_j$ computes a finite-difference approximation of the gradient at each pixel $j$,
and the 2-norm also acts pixel-wise on the differences. 
In CT, for example, the typical image consists of regions having 
an approximately 
constant gray-level value separated
by sharp boundaries between various tissue types. The magnitude of the spatial gradient of such
images is zero within constant regions and non-zero along edges, so the gradient magnitude image can be sparse.
The $\ell_1$-norm applied to the gradient magnitude image is known as the total variation (TV) semi-norm, 
\begin{equation}
\label{eq:TV}
\TV{\vec{f}} = \| \Psi (\vec{f})\|_1 = \sum_{j=1}^{N_\text{pix}} \|D_j\vec{f}\|_2,
\end{equation}
and the optimization problem of interest becomes
\begin{equation}
\label{TV}
\vec{f}^* = \argmin_{\vec{f}} \; \| \vec{f} \|_{TV} \text{ such that } X \vec{f} = \vec{g}.
\end{equation}

However, the theoretical results from CS do not extend to the CT setting. 
Three properties that separate CT matrices from typical CS matrices are that CT matrices
\begin{enumerate}
 \item are structured and do not belong to random matrix classes
for which CS results are proved \cite{candes2008introduction},
 \item can have rank smaller than the number of rows,
which means that there exist vectors $\vec{g}$ in the data space that
are inconsistent with $X$, and accordingly the linear imaging model  \eqref{xfeg} has no solution, and
\item may be numerically ill-conditioned in case of having more rows than columns (data set size is greater than the image representation).
\end{enumerate}
Nevertheless, it has been demonstrated empirically
in extensive numerical studies with computer phantoms under ideal data conditions as well as with actual scanner data
that highly accurate reconstructed images for ``undersampled''
projection data can be obtained from \eqref{L1} and variants thereof. 

It is precisely this last phrase which is of interest in the present paper:
 what exactly does it mean to have ``undersampled'' data
for CT? Under-sampled data implicitly relies on a certain level of
 sampling being sufficient---but no such precise concept exists for CT using the DD imaging model,
to the best of our knowledge. Without a reference point for having sufficient sampling
it is difficult to quantify admissible levels of undersampling. 
In the present paper, we aim to provide this reference point.
Specifically in the following section, we propose sufficient-sampling conditions (SSCs)
to be computed for specific system matrix classes, and
which serve as a reference for quantifying the admissible undersampling for
reconstruction with sparsity-exploiting methods.
The application of the SSCs is demonstrated with numerical simulations 
of breast CT.

\section{Sufficient-sampling conditions}
\label{sec:svd}

In considering sufficient sampling for circular fan-beam CT, the CC model is recast
as a CD model by introducing a discrete sampling operator,
usually taken to be evenly distributed delta functions, on the
CT sinogram space.  Making the assumption that the underlying 
sinogram 
function is band-limited,
many useful and widely applicable results have been obtained, see for example Section 3.3 of
\cite{natterer1986mathematics} and Refs. \cite{natterer:1993,izen:2005,faridani:2006}. Furthermore, for more 
advanced scanning geometries and sampling patterns there are available tools for analysis
such as singular value decomposition (SVD), direct analysis of multi-dimensional aliasing, and
the evaluation of the Fourier crosstalk matrix \cite{barrett:94,barrett:95,lariviere:2004}.
These important results, however, do not apply directly to IIR, since for the DD model we need to take into account the finite image expansion set.

We consider an empirical approach for characterizing sufficient sampling within a class of system matrices. The idea is to 
fix the image representation, which for the circular fan-beam 
system matrix 
class is the parameter $N_\text{pix}$, 
and then 
vary the sampling parameters $\nv$ and $\nb$ 
to ensure accurate determination of the
pixel values.
This is done by establishing sufficient-sampling conditions (SSCs) based on matrix properties in the considered system matrix class.
If the 
system matrix
class is altered, the SSC-analysis must be
redone.


\subsection{SSC definitions} \label{sec:ssc_def}
The SSCs, we propose, characterize invertibility and ill-conditioning of the 
system matrix 
class.
In considering the DD imaging
model \eqref{xfeg}, the data $\vec g$ are restricted to the range of $X$.
This separates out the issue of model inconsistency which does not
have direct bearing on sampling conditions.

We define SSC1 to be sampling such that $X$ has at least as many rows as columns. 
If there are fewer rows than than columns, $X$ necessarily has
a non-trivial null space, and solutions of \eqref{xfeg} will not be unique.
Even if the number of rows is equal to or larger than the number of columns,
there may still be a non-trivial null space, because the rows can be linearly dependent. 
In addition to SSC1, we define SSC2 to mean that $X$ has a null space consisting only of
the zero image, or equivalently, that the smallest singular value $\sigma_\text{min}$ of $X$ is non-zero.
Existence of a unique solution to \eqref{xfeg} is ensured by SSC2.
Both SSC1 and SSC2 can be
evaluated for any system matrix 
class.

Neither of SSC1 and SSC2 address numerical instability and to address that,
we employ the condition number of $X$,
the ratio of the largest and smallest singular values,
\begin{equation}
\label{cn}
\kappa(X) = \svalmax / \svalmin.
\end{equation}
The condition number $\kappa$ can be as small as $1$ and
the larger $\kappa$ becomes, the more numerically unstable is solution of $X\vec{f} = \vec{g}$.
How to use $\kappa$ to define a SSC requires some discussion.

Whereas sensing matrix classes studied in CS typically are well-conditioned---for instance the square discrete Fourier transform (DFT) matrix is orthogonal, thus having a condition number of $1$---the system matrices encountered in
X-ray CT can have a relatively large condition number \cite{natterer1986mathematics,huesman1977effects}, 
which leads to numerical instability and thus large sensitivity to noisy measurements. 
Even if SSC2 holds, the condition number $\kappa$ is finite but may still be large, potentially allowing
other images than the desired solution to be numerically close to satisfying $X\vec{f} = \vec{g}$.
If we fix the image representation, which for the present 2D circular fan-beam setup
amounts to fixing $N_\text{pix}$, and increase the sampling, allowing 
$N_\text{views}$ and $N_\text{bins}$ to increase toward $\infty$,
the condition number will decrease toward a limiting condition number,
\begin{equation}
\label{kappaDC}
\kappa_{DC} = \lim_{N_\text{views},N_\text{bins} \rightarrow \infty} \kappa (X),
\end{equation}
where the $DC$ subscript refers to the fact that $X$ is limiting to a discrete-to-continuous (DC) system
matrix. The limiting
condition number $\kappa_{DC}$ is the best-case $\kappa$ for a fixed image representation,
but $\kappa_{DC}$ may 
still 
be larger than $1$.

For actual CT scanners, 
it is not practical to allow $N_\text{views}$ and $N_\text{bins}$ to increase without bound,
and empirical experience shows diminishing improvements in doing this.
To balance the impracticality of going to continuous sampling on the one hand 
against the need to optimize numerical stability on the other,
we introduce SSC3 to mean that the condition number of $X$ satisfies 
\begin{equation}
 \kappa (X)/ \kappa_{DC} < r_\text{samp},
\end{equation}
where $r_\text{samp}$ is a finite ratio parameter greater than $1$.
The smaller the choice of $r_\text{samp}$, the closer $X$ is to the DC limit.
This SSC can also be generally applied to other 
system matrix 
classes, but the appropriate
parameter setting of $r_\text{samp}$ will be specific to a particular class. 

Finally, we introduce SSC4 specifically for the present
2D circular fan-beam 
system matrix class. This SSC is taken to mean $2N$ samples in both
the view and bin directions, i.e., $N_\text{views}=N_\text{bins}=2N$. This SSC is simple
to evaluate, and we will demonstrate empirically that it is a useful condition,
which acts as a good approximation for attaining SSC3 with $r_\text{samp} = 1.5$.
This SSC is specific to the 
system matrix class investigated here. Even slightly different system matrix classes might not allow for the same SSC4 definition.

Our strategy is similar to analysis presented in early works on CT, such as in \cite{huesman1977effects},
but the point here is not novelty of the analysis;
rather we need to establish a reference point by which
to evaluate the sampling reduction admitted by sparsity-exploiting methods.

In what follows, the proposed SSCs are examined for the 2D circular fan-beam 
system matrix class. First, small systems are considered, where $X$ can be explicitly
computed and analyzed so that the full set of singular values of $X$ is attainable.
Second, we argue that our conclusions generalize to larger, more realistic systems,
where $X$ is impractical to store in computer memory and it is only feasible to
compute the smallest and largest singular values.

\subsection{SSCs for small systems}
\label{sec:smallsys}

\begin{figure*}[!t]
\begin{minipage}[b]{0.5\linewidth}
\centering
\centerline{\includegraphics[width=\linewidth]{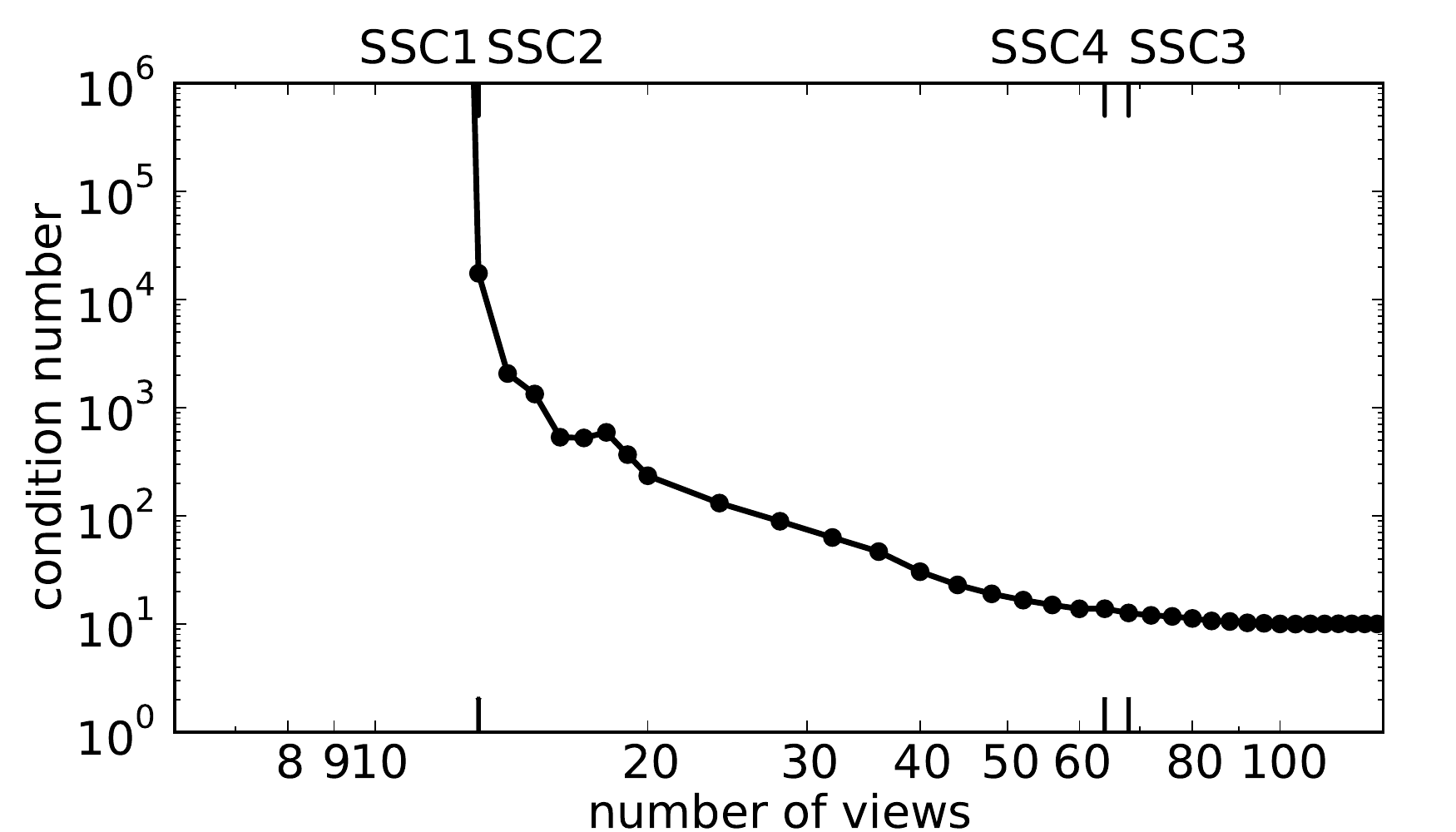}}
\end{minipage}
\begin{minipage}[b]{0.5\linewidth}
\centering
\centerline{\includegraphics[width=\linewidth]{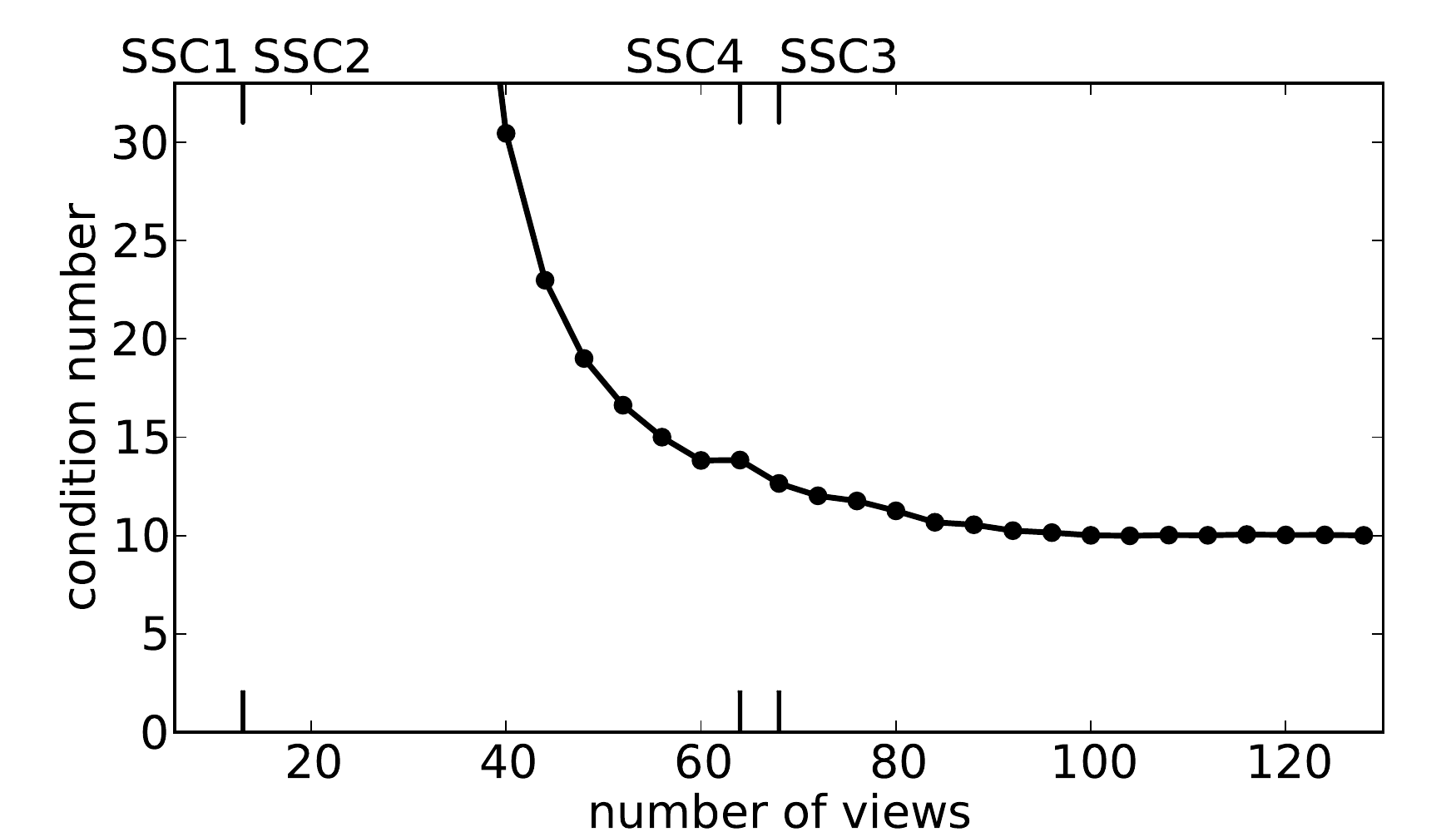}}
\end{minipage}
 
\begin{minipage}[b]{0.5\linewidth}
\centering
\centerline{\includegraphics[width=\linewidth]{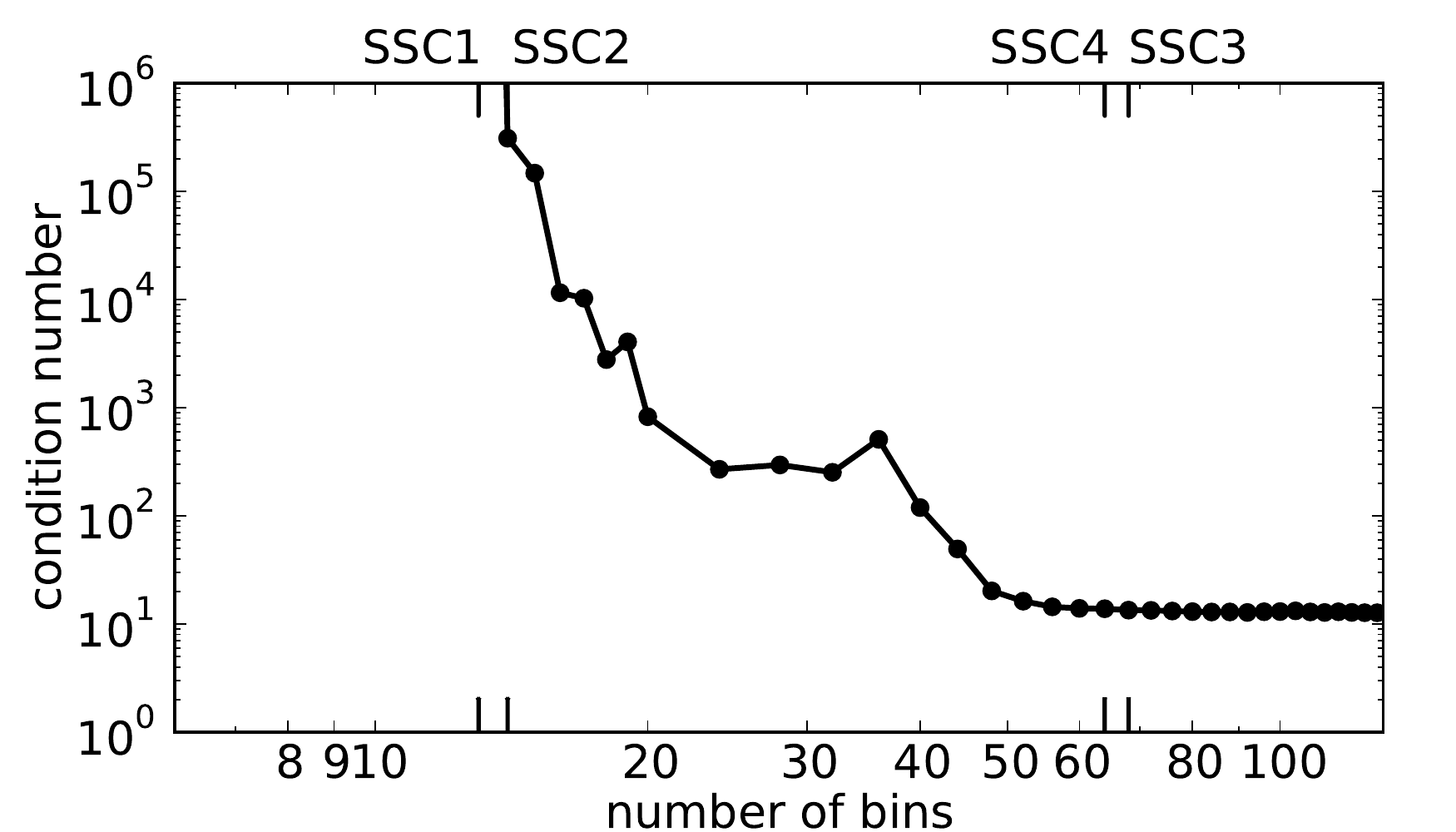}}
\end{minipage}
\begin{minipage}[b]{0.5\linewidth}
\centering
\centerline{\includegraphics[width=\linewidth]{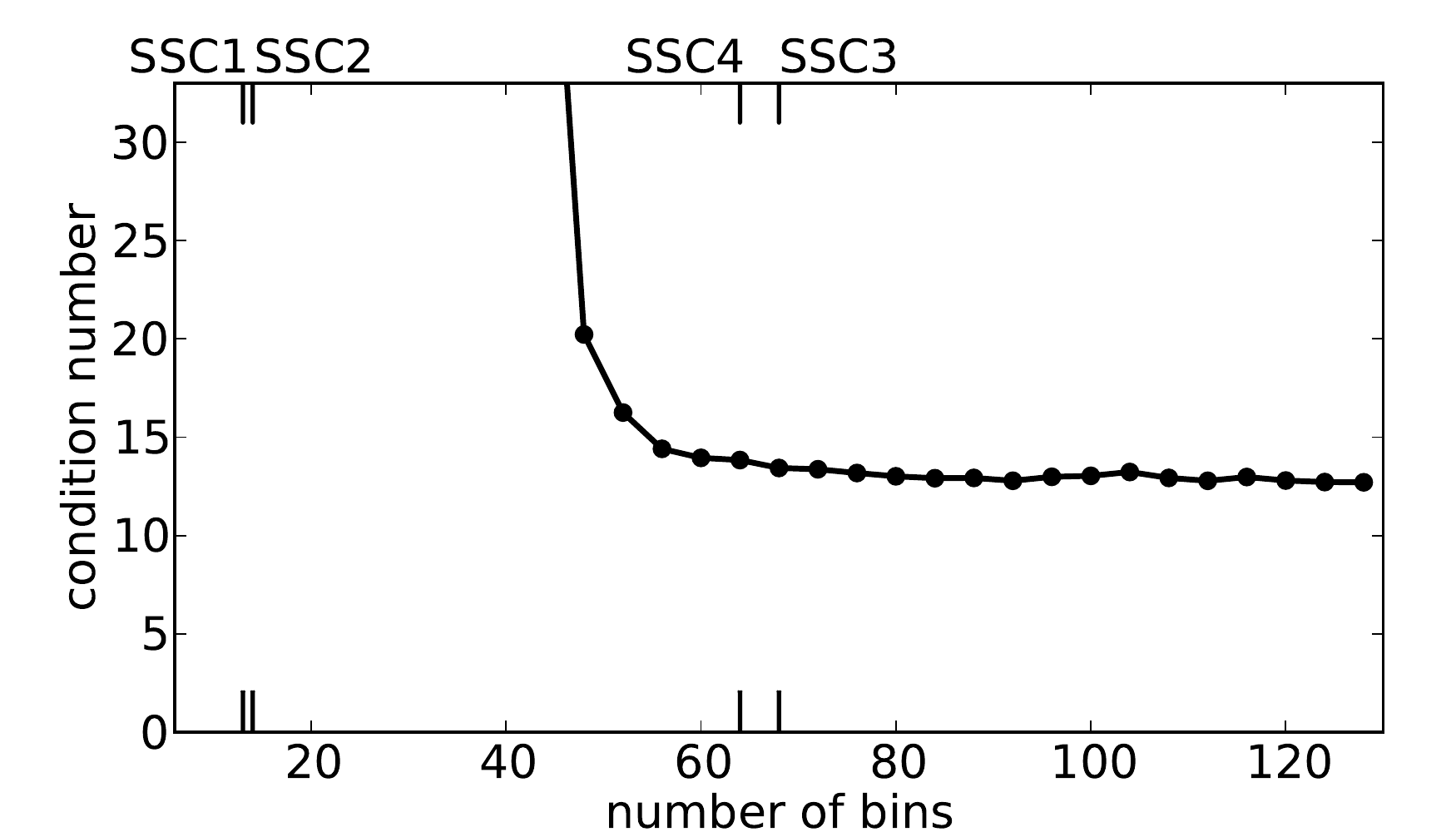}}
\end{minipage}
\caption{
Condition numbers for system matrices (line-intersection) modeling circular fan-beam projection
data from the $812$-pixel circular FOV contained within a $32 \times 32$ pixel square.
Top: number of bins is fixed at $64$. Bottom: the number of views is fixed at $64$. Left:
Double-logarithmic plot for overview. Right: Linear plot of details.
The abbreviations SSC1, SSC2, SSC3 and SSC4 are the sufficient-sampling conditions
discussed in Sec. \ref{sec:ssc_def}. 
\label{fig:cond_views_bins}}
\end{figure*}

We consider a small $N=32$ image array with 
$N_\text{pix} = 812$, 
and generate system matrices $X$ for different
numbers of views, $N_\text{views} \in [8,128]$, and detector bins, $N_\text{bins} \in [8,128]$. 
The condition number $\kappa (X)$ is computed
through direct SVD of $X$ for all values of $N_\text{views}$ and $N_\text{bins}$
within the specified parameter ranges, 
and $\kappa(X)$ as a function of $N_\text{views}$ (for fixed $N_\text{bins} = 64$) 
and $N_\text{bins}$ (for fixed $N_\text{views} = 64$) is shown in Fig. 
\ref{fig:cond_views_bins}.
The plots show three phases:
the left-most part, where the condition number is infinite;
the middle, where the condition number becomes finite and decays slowly;
and the right-most part, where it remains relatively stable.
The positions of the different SSCs are shown at the top and bottom,
and they serve as transition points between the three phases. 

For varying the number of views, SSC1 occurs at
$N_\text{views} = 13$, where
$X$ is of size $832$ by $812$. In fact, also
SSC2 occurs here, since $\kappa$ has become finite. 
For varying the number of bins, SSC1 occurs at the same place, but
SSC2 needs $14$ bins.
In general, we have no way to determine whether SSC1 and SSC2 occur in the same position for the whole
system matrix class, which makes SSC1 less reliable as a general reference of full sampling.
On the other hand, SSC2 is a reliable reference point for full sampling, however, 
SSC2 requires more work to determine, because its location can change with a change of system matrix class.

After passing SSC2, the condition number $\kappa$ decreases.
For larger $\nv$ and $\nb$, the decay becomes slower, and we pick 
$r_\text{samp} = 1.5$ as a trade-off between a
sufficiently small condition number and a finite number of views.
As an approximation of $\kappa_{DC}$ we take the value of $\kappa$
at $N_\text{views} = N_\text{bins}= 4N = 128$, yielding $\kappa_{DC} = 9.17$. 
Then SSC3 occurs at $\nv = 68$ and at $\nb = 68$,
which suggests a symmetry in $\nv$ and $\nb$.
On the other hand, the decrease in $\kappa$ during the middle part is not symmetric in the parameters
$N_\text{views}$ and $N_\text{bins}$; the decrease in $\kappa$ with $N_\text{views}$
is gradual while that of $N_\text{bins}$ is step-like at $N_\text{bins}=48$.
Nevertheless, at the position of SSC3, there is only small further reductions
in $\kappa$ to be gained by going to larger $\nv$ and $\nb$. The simpler condition
SSC4 occurs at $\nv = 64$ and $\nb = 64$, and it
approximates SSC3 with $r_\text{samp} = 1.5$ closely.

\subsection{Altering the system matrix class} \label{sec:ssc_alternate_model}
\begin{figure*}[!t]
\begin{minipage}[b]{0.5\linewidth}
\centering
\centerline{\includegraphics[width=\linewidth]{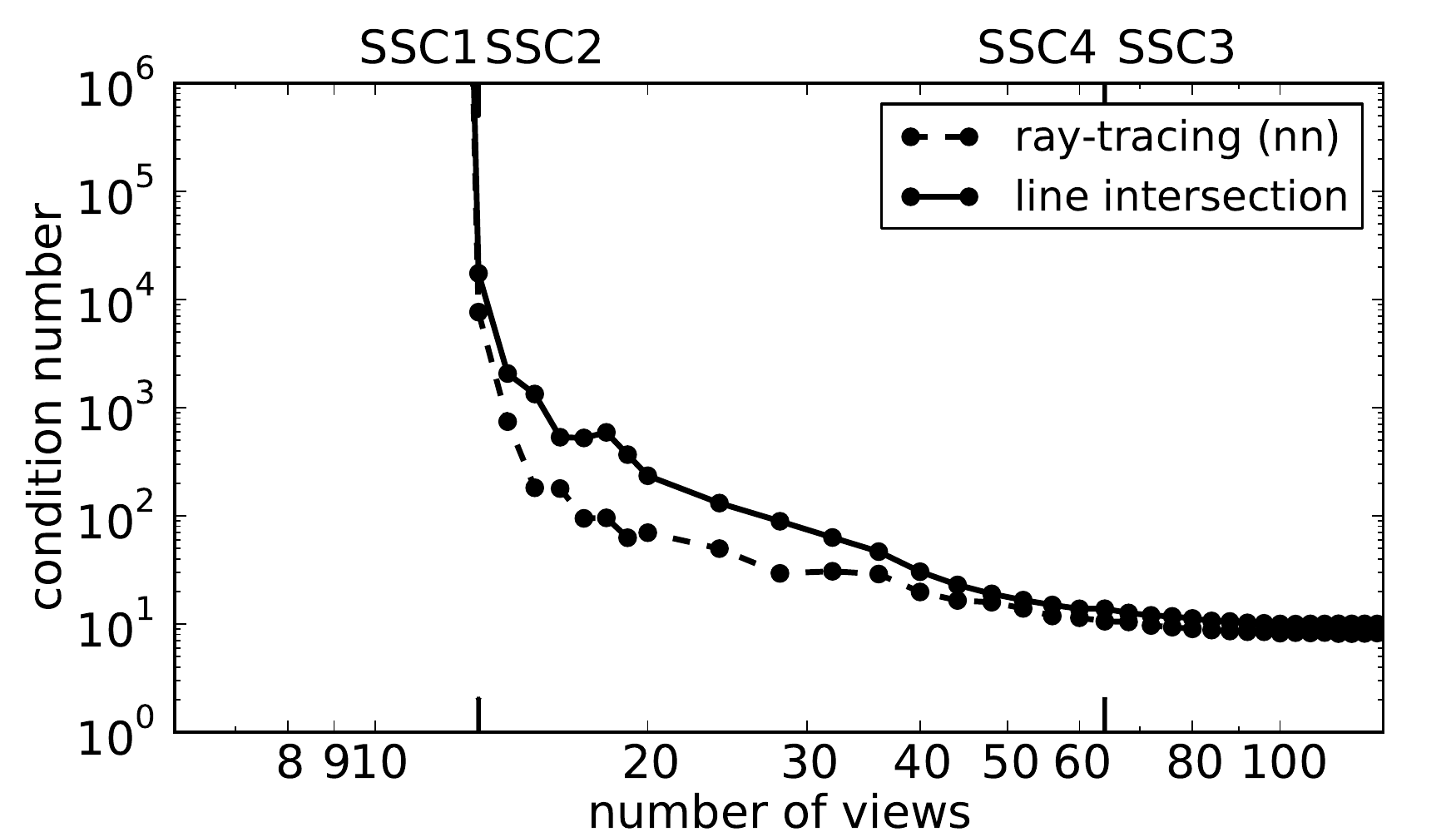}}
\end{minipage}
\begin{minipage}[b]{0.5\linewidth}
\centering
\centerline{\includegraphics[width=\linewidth]{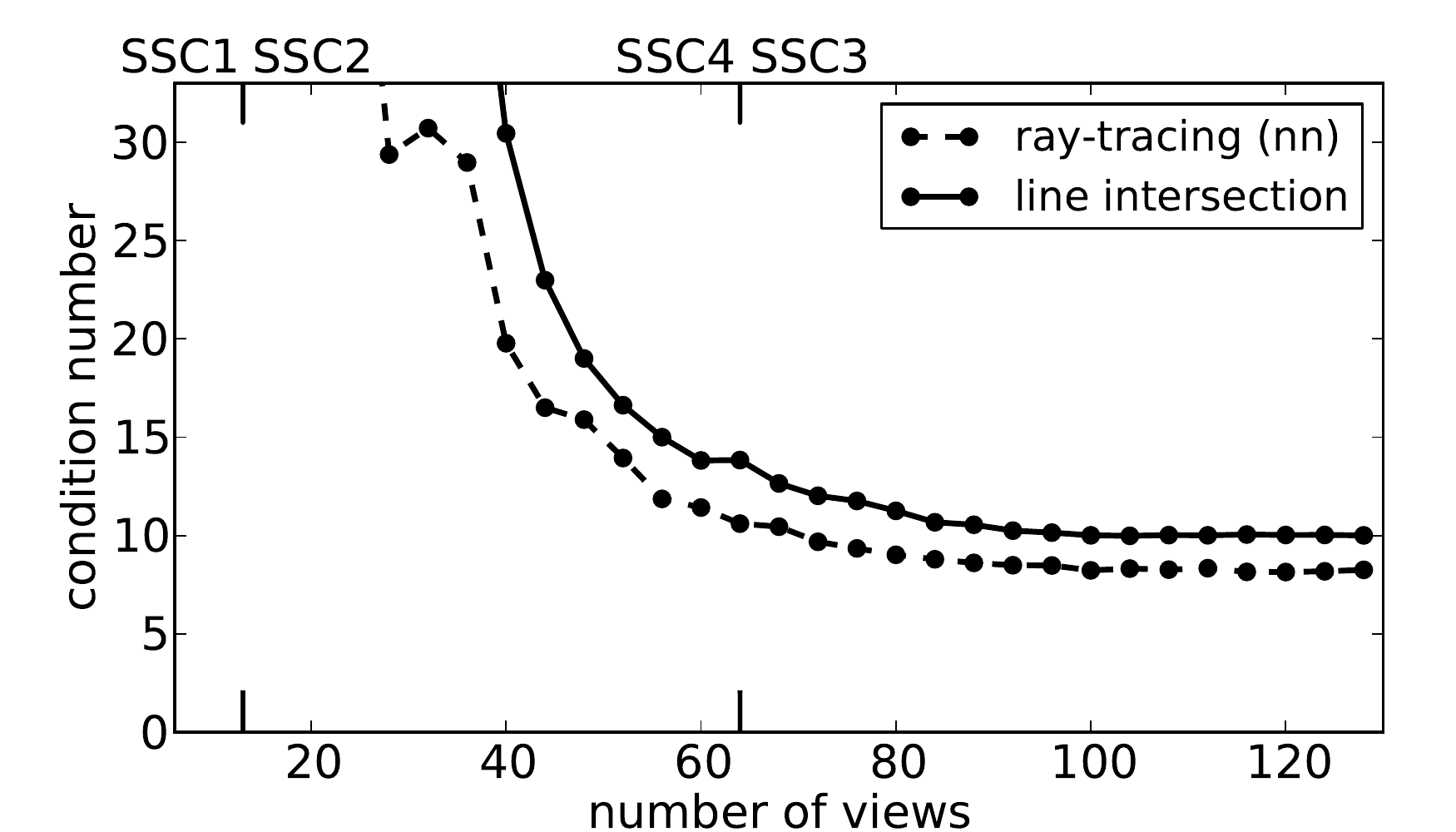}}
\end{minipage}
 
\begin{minipage}[b]{0.5\linewidth}
\centering
\centerline{\includegraphics[width=\linewidth]{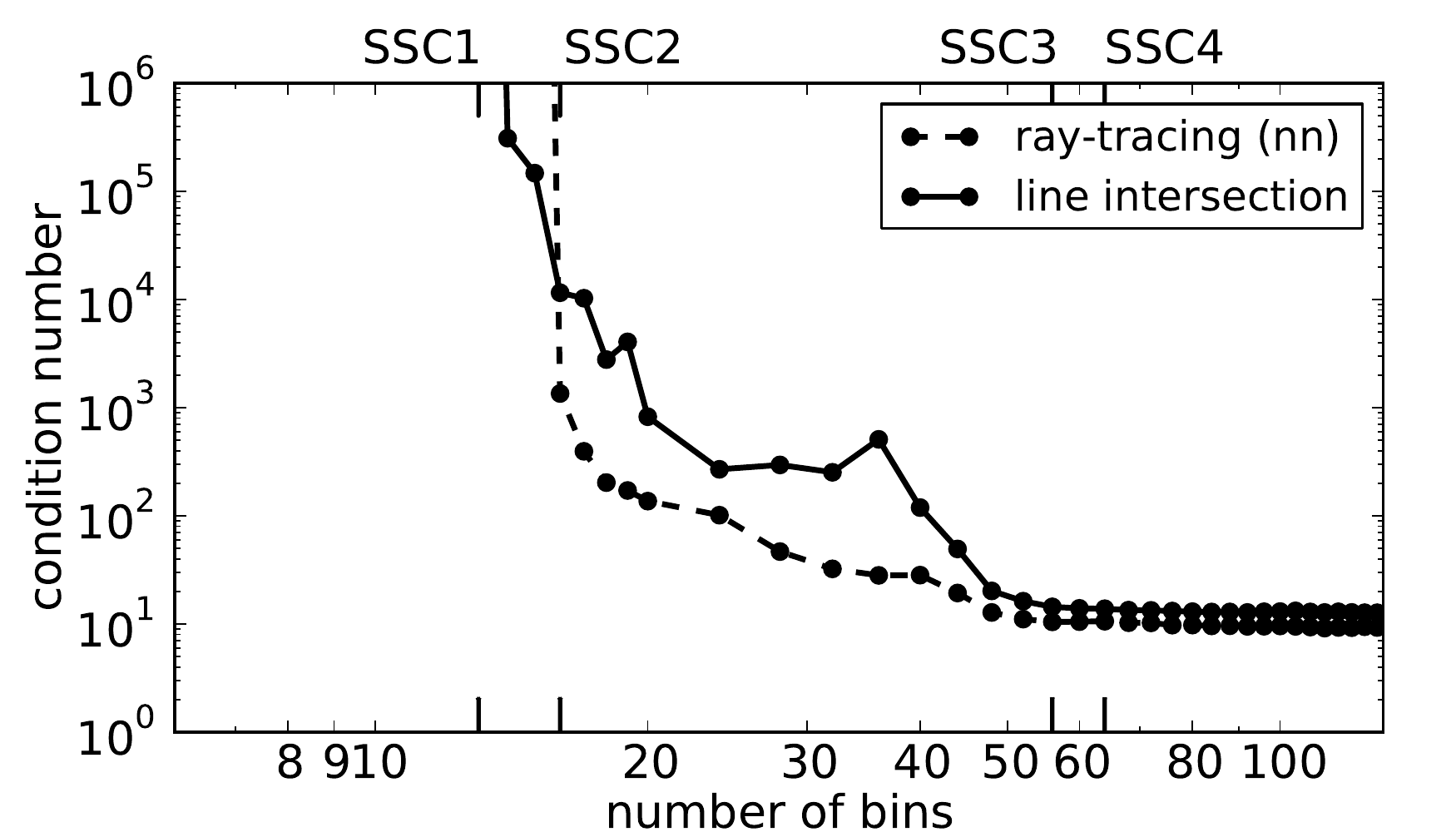}}
\end{minipage}
\begin{minipage}[b]{0.5\linewidth}
\centering
\centerline{\includegraphics[width=\linewidth]{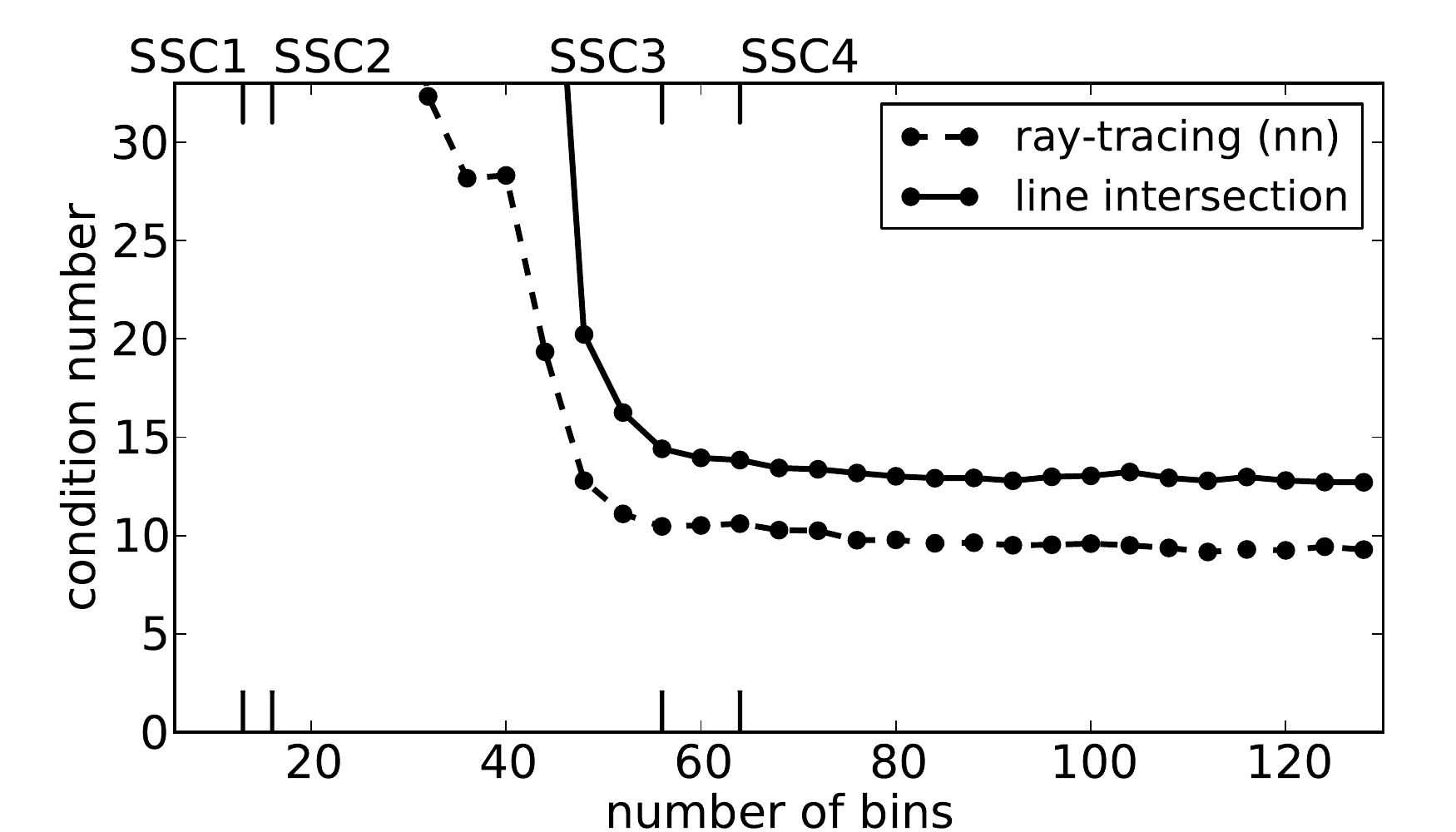}}
\end{minipage}
\caption{
Condition numbers for system matrices (ray-tracing with nearest
neighbor interpolation, and line-intersection for comparison) modeling circular fan-beam projection
data from the $812$ pixels circular FOV contained within a $32 \times 32$ pixel square.
Top: number of bins is fixed at $64$. Bottom: the number of views is fixed at $64$.
Left: double-logarithmic plot for overview. Right: linear plot of details. The labels
SSC1, SSC2, SSC3 and SSC4 are the sufficient-sampling conditions, discussed in
Sec. \ref{sec:ssc_def}, for the ray-tracing system matrix. 
\label{fig:modelComp}}
\end{figure*}

Altering the system matrix 
class will in general alter the SSCs.
To demonstrate this effect, we replace the line-intersection based system matrix class
by ray-tracing, using nearest-neighbor interpolation at the mid-line of each pixel row.
The experiment is repeated and the obtained condition numbers are shown in
Fig. \ref{fig:modelComp}, along with the ones based on line-intersection, for comparison.
The shown SSCs are for ray-tracing. While the same overall trends are seen,
there are some significant differences.
First, for fixed $\nv = 64$ we need $\nb = 16$ to obtain SSC2, compared to $14$
for line-intersection. This firmly establishes that SSC1 does not imply SSC2,
and that the precise position of SSC2 cannot be inferred from knowing SSC2 of a similar 
system matrix.
Second, the ray-tracing condition numbers are smaller than the same for
line-intersection, for instance, $k_{DC}=7.23$ is $20\%$ lower relative to
the line-intersection version of $X$. That the ray-tracing condition numbers are lower
does not necessarily mean that this method is ``better'' than the line-intersection
method for real-world applications, because the other side of the story is model error,
which is not considered here.
Finally, the positions of SSC3 are different, for fixed $\nb = 64$ coinciding with
SSC4, while for fixed $\nv = 64$ occurring at $\nb = 56$.
Still, for larger $\nv$ and $\nb$ there are only small further reductions in $\kappa$ to be gained,
and SSC4, at $r_\text{samp} = 1.45$, approximates SSC3 closely.

One could imagine that other system matrix classes 
such as employing area-weighted integration instead of the linear
integration or different basis functions could alter the condition numbers of $X$ even more
substantially. We do not include results for more system matrix classes,
as our goal here is not to provide a comprehensive comparison between all
conceivable classes, but merely to stress that different classes can 
have different SSCs, and to propose carrying out the same study for gaining insight in
the particular 
system matrix class at hand.

\subsection{SSCs for larger systems} \label{sec:ssc_larger}
\begin{figure*}[!t]
\begin{minipage}[b]{0.5\linewidth}
\centering
\centerline{\includegraphics[width=\linewidth]{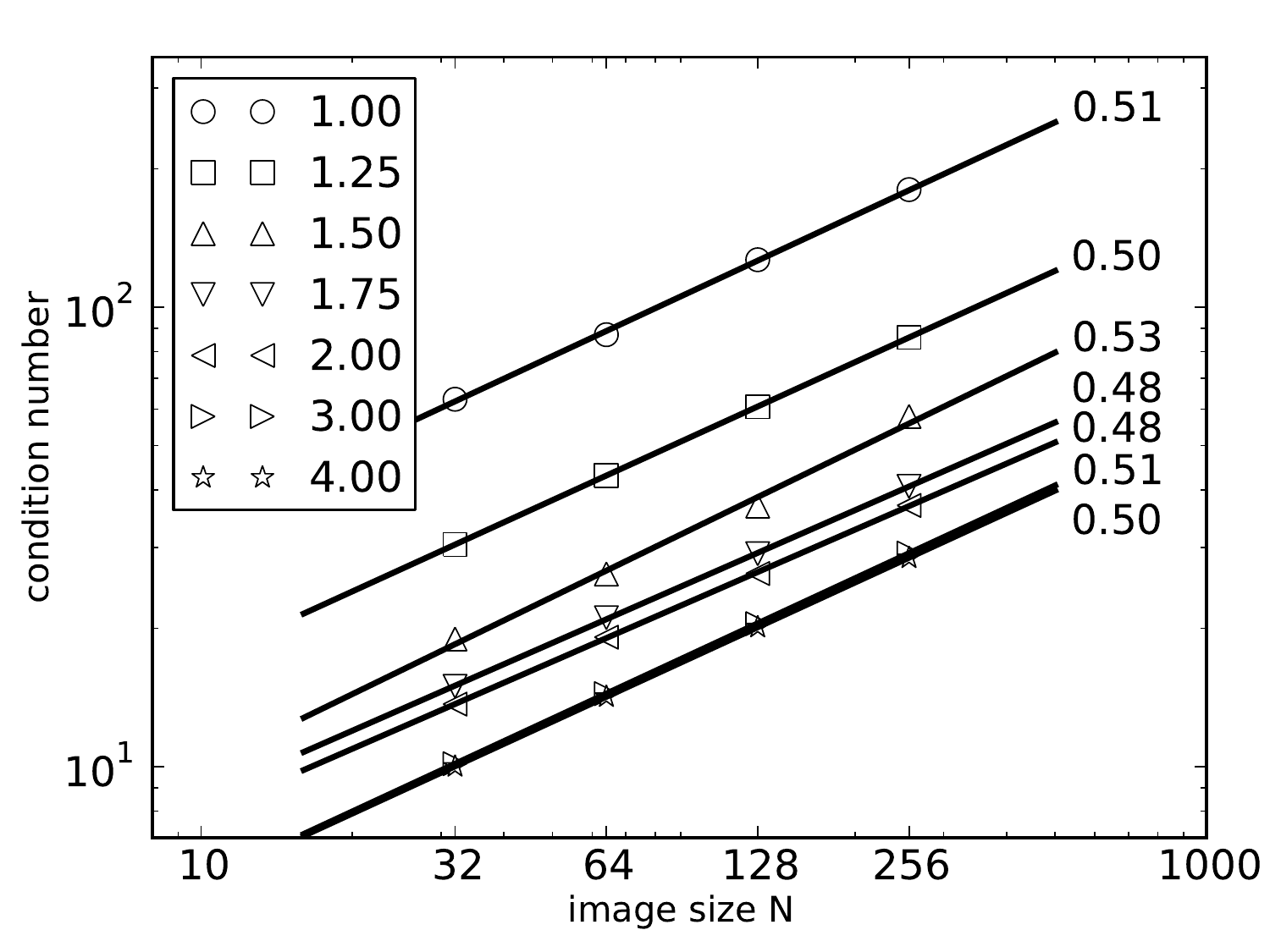}}
\end{minipage}
\begin{minipage}[b]{0.5\linewidth}
\centering
\centerline{\includegraphics[width=\linewidth]{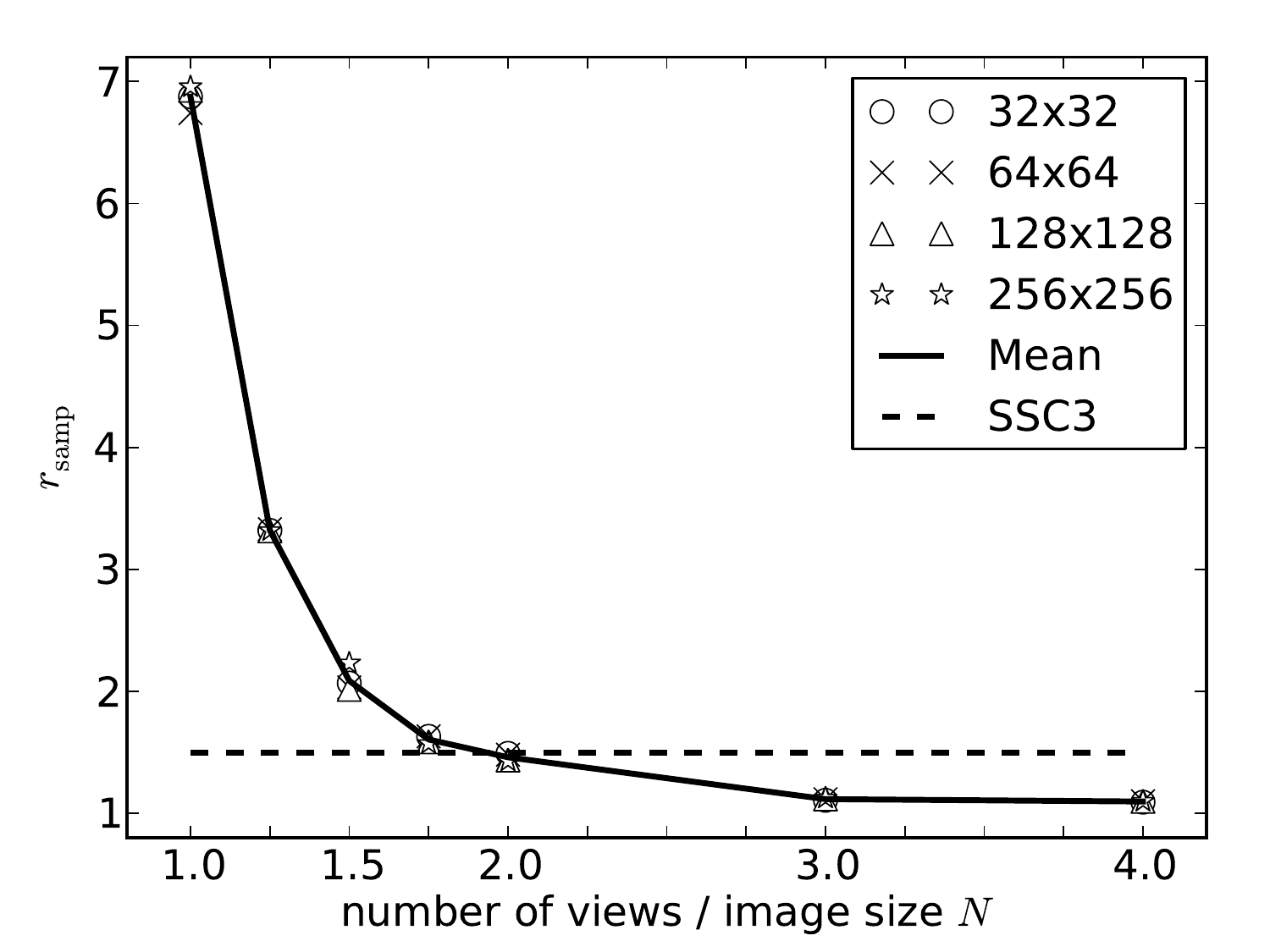}}
\end{minipage}
\caption{Left: condition numbers as function of image size.
Each symbol represents different number of views ranging from $N$ to $4N$.
Circular image arrays are used with sizes given by 
$N=32$, $64$, $128$, $256$. The number
of bins is fixed at $2N$.
With each number of views is also shown the best linear fit and its slope is given.
In all cases the condition number scales with $\sqrt{N}$.
Right: same condition numbers normalized by the
respective $\kappa_\text{DC}$ at each $N$ and plotted as function of
view number normalized by image size $N$.
The full line is the point-wise mean and the dashed line
is the position of SSC3 with $r_\text{samp} = 1.5$.
Independently of $N$, SSC3 with $r_\text{samp} = 1.5$ occurs very close to $\nv = 2N$.
\label{fig:svdScaling}}
\end{figure*}
The results shown in Fig. 
\ref{fig:cond_views_bins} and Fig. 
\ref{fig:modelComp}
give a sense about various sampling combinations, but
the system size is unrealistically small.
In this section we aim to extend the results to larger systems.
For large $X$, it is not practical to compute the direct SVD for evaluating $\kappa$.
Instead, we seek only 
to obtain $\svalmin$ and $\svalmax$, which can be accomplished
through the power and inverse power methods \cite{Saad:1992}.

For characterizing the present circular, fan-beam 
system matrix 
class,
$\kappa(X)$ is computed for larger image arrays 
with sizes $N = 32$, $64$, $128$, $256$.
We focus on sampling conditions
where $N_\text{bins} = 2N$ and report the data sampling in views, $N_\text{views}$,
as multiples of $N$, ranging from $1.0$ to $4.0$.
The left plot in Fig. \ref{fig:svdScaling} shows the condition number as a function
of $N$ for each sampling size on a double logarithmic scale. A clear linear trend
is seen in all cases and the best linear fits and their slopes are also shown,
in all cases very close to $0.50$, and we conclude that $\kappa$ scales with $\sqrt{N}$.
For increasing $\nv$, the condition numbers tend towards the bottom line,
note in particular that not much difference is seen between $\nv = 3N$ and $\nv = 4N$
indicating that the limiting $\kappa$ is approached.
We conclude that $\kappa_{DC}$ also scales with $\sqrt{N}$.
As a result, it can expected that at SSC4,
i.e., $\nv = \nb = 2N$, $\kappa(X)/\kappa_{DC} \approx 1.5$, which was the case at $N=32$.
Hence, SSC4 will continue to approximate SSC3 closely, when the image size is increased.
To further support this conclusion we show in the right plot of Fig. \ref{fig:svdScaling}
the ratio $r_\text{samp} = \kappa(X)/\kappa_{DC}$ as function of the number of views
(normalized by $N$) for each $N$. The $r_\text{samp}$
values are almost identical for all $N$ and intersect the line $r_\text{samp} = 1.5$
very close to $\nv = 2N$, which is precisely SSC4.

\subsection{Summary of SSCs}

The conditions SSC1 and SSC2 are useful reference points for invertibility of $X$ and can be computed for 
any system matrix class.
The size of the gap between SSC1, $X$ being square, and SSC2, $X$ having an empty null space,
is governed 
by inherent linear dependence of the rows
of the system matrix. 
Because the results show little
difference between SSC1 and SSC2 for the present 
system matrix 
class and SSC1 is easier to compute, we use only SSC1 in the simulation studies in the following section.

For 
system matrix 
classes representing CT imaging, stability of the system matrix plays an important role, and
accordingly we have introduced SSC3 which also can be computed for
any 
system matrix 
class. For the present, circular fan-beam 
system matrix 
class, SSC3 at $r_\text{samp} = 1.5$ is a useful operating point, and this level of sampling is well approximated by the
simple rule, SSC4, where $\nv=\nb=2N$.  We point out that for other
system matrix 
classes, even those representing circular fan-beam CT, other operating
points for SSC3 may be more appropriate and empirical studies must be
performed to see if a simple condition, such as SSC4, can approximate accurately SSC3.

In the remaining part of the paper we demonstrate 
how we can use the SSCs as a reference for stating admissible undersampling
factors in sparsity-exploiting reconstruction.

\section{Numerical experiments with SSCs and sparsity-exploiting undersampled reconstruction}
\label{sec:numerical}
In this section we investigate sparsity-exploiting IIR in numerical simulation studies. 
Our goal is to numerically demonstrate and quantify the undersampling admitted by sparsity-exploiting IIR, i.e., at which an accurate reconstruction is obtained.
We use numerical simulation,
because we are unaware of any theoretical 
results establishing undersampling guarantees for the present system matrix
class.
%
We focus here on exploiting
gradient-magnitude sparsity by use of constrained TV-minimization.

Three important factors differentiate
the present studies from previous simulation work with constrained TV-minimization:
\begin{enumerate}
 \item use of phantoms with realistic complexity, 
 \item numerically accurate solution to the constrained TV-minimization problem, and
 \item quantitative references for full sampling---the central topic of the paper.
\end{enumerate}
For each factor, we we briefly discuss the significance.

Much simulation work on constrained TV-minimization has used regular, piece-wise
constant phantoms, such as the Shepp-Logan phantom, 
to demonstrate the promise of the technique. For that purpose, 
such unrealistically simple phantoms were fine, and simulations were generally followed up by demonstration with
actual CT projection data. For the present purpose of quantifying admissible undersampling, we need
phantoms with similar complexity as would be encountered
in CT applications, and as an example we focus on breast CT.
The standard measure of complexity employed in CS is the image sparsity, i.e., the number of non-zeroes in the image, or in the case of TV-minimization the number of non-zeroes in the gradient-magnitude image.
Accordingly, we choose a digital phantom with realistic gradient-magnitude sparsity 
modeling breast anatomic structure \cite{Reiser2010}.

The accuracy requirement on the solver of constrained TV-minimization for the present
study is extremely high. The optimization problems in Sec. \ref{sec:setup}, below, are
solved to high accuracy, which has been made possible only recently for large-scale CT
problems involving the TV-semi-norm through development of advanced
first-order methods \cite{Jensen2012,ChambollePock:2011,Sidky2012}.
This level of accuracy is necessary, because empirical image error results obtained by sweeping parameters of the 
system matrix 
class will be used for 
determining whether a numerically computed solution is close to the original.
High-accuracy solutions
remove any doubt about whether the resulting images, and the corresponding quantitative
measures, depend on 
the algorithm used to solve constrained TV-minimization.

The SSCs defined above provide reference points useful for interpreting the empirical
results of this section and help to quantify undersampling admitted by constrained TV-minimization.

\subsection{Breast CT background}
\begin{figure*}[!t]
\begin{minipage}[b]{\linewidth}
\centering
\centerline{\includegraphics[width=\linewidth]{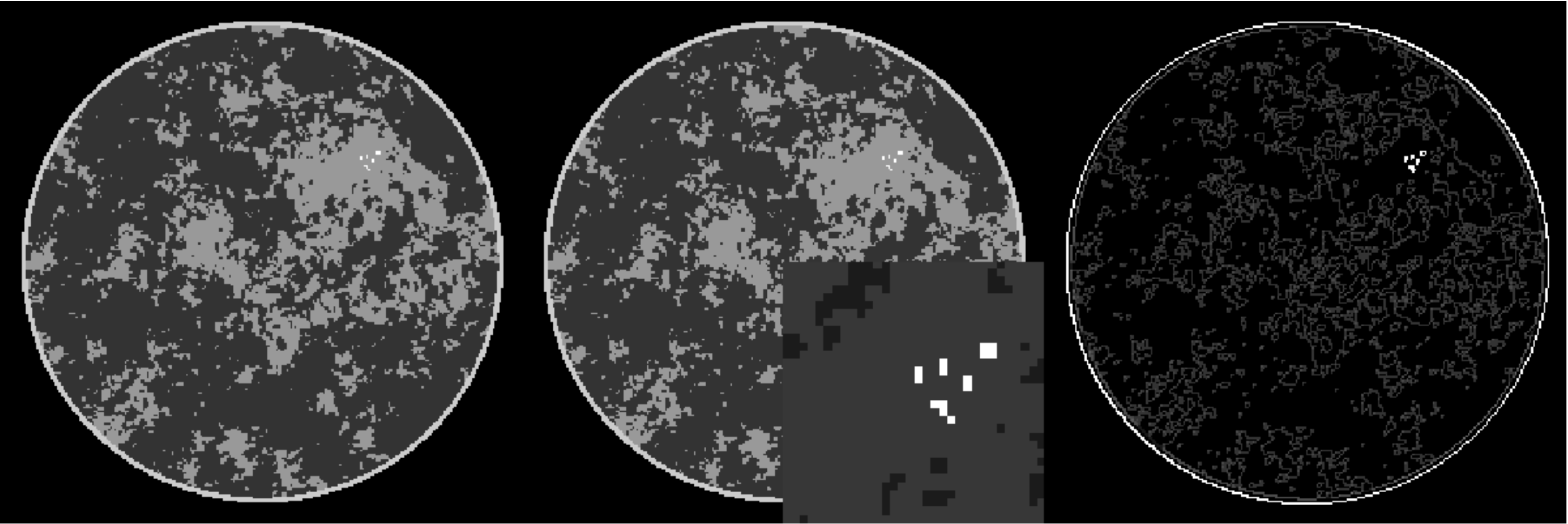}}
\end{minipage}
\caption{Left: $256 \times 256$ pixelized breast CT phantom used in the
present study. 
Middle: same with ROI around microcalcifications shown magnified as inset.
Right: the gradient magnitude image, which has a sparsity of $10,019$
non-zero pixel values.
\label{fig:breastPhantom}}
\end{figure*}

Breast CT  \cite{boone2001dedicated,glick2007breast} is being considered as a possible screening or diagnostic tool for breast cancer.
The system requirements are challenging from an engineering standpoint, because this type
of CT must operate with a total exposure similar to two full-field digital mammograms (FFDM).
FFDMs for a screening exam entail two X-ray projections, while breast CT acquires on the order
of 500 X-ray projections. The exposure previously used for only two views is now divided
up among 250 times more projections. Accordingly, sparsity-exploiting IIR algorithms for CT 
may have an impact on the breast CT application. The potential
to reconstruct volumes from fewer views
than a typical CT scan might allow an increased exposure per view. 

For the present study, we employ the breast phantom originally described in \cite{Reiser2010}
and displayed in Fig. \ref{fig:breastPhantom}.
It consists of $N_\text{pix} = 51,468$ pixels within the circular image region,
contained in a $256 \times 256$ array.
The breast phantom
has a small region of interest (ROI) containing 5 tiny ellipses which model microcalcifications. 
The gray values
range from 1.0 to 2.3.
The modeled tissues and corresponding
gray values are fat at 1.0, fibroglandular tissue at 1.10, skin at 1.15, and microcalcifications
ranging from 1.9 to 2.3. 
The sparsity in the gradient magnitude image is 
$10,019$, or roughly one fifth
of $N_\text{pix}$. Because we are investigating the utility of gradient-magnitude sparsity-exploiting
algorithms, it is important that the test phantom have a realistic sparsity level relative to the
actual application.

\subsection{Simulation optimization problems and algorithms}
\label{sec:setup}
Our goal is to evaluate quantitatively what level of undersampling
reconstruction through \eqref{TV} allows. 
Similar to the analysis of the linear imaging model \eqref{xfeg},
only data $\vec g$ in the range of $X$ is considered.
Although not a realistic assumption for actual CT data, this ``inverse crime''
scenario \cite{Kaipio2005} is appropriate for obtaining a reference of the
underlying admissible undersampling.
For the numerical studies, we solve a relaxed form of \eqref{TV}, where the data equality
constraint is replaced by an inequality
allowing for a small
deviation $\epsilon$ from data as measured by the distance $D$
between the data $\vec g$ and the projection $X\vec f$ of some image $\vec f$,
\begin{equation}
 D(X\vec f, \vec g) \leq \epsilon, \label{eq:constraint}
\end{equation}
where
\begin{equation*}
D(X\vec f, \vec g)^2 = \frac{1}{N_\text{views} N_\text{bins} } \|X\vec{f} - \vec{g}\|_2^2.
\end{equation*}
Scaling the data error $D$
with $\nv$ and $\nb$ is done to enable comparison across images reconstructed
from different view and detector-bin numbers.
The constrained TV-minimization problem is
\begin{align}
\text{\ltwotv{}:} \qquad &\vec{f}^* = \argmin_{\vec{f}} \| \vec{f} \|_\text{TV} \label{eq:ltwotv} \\
 \qquad \text{subject to} \qquad &D(X\vec f, \vec g) \leq \epsilon. \notag
\end{align}

Accurate solution of
\eqref{eq:ltwotv} is non-trivial; although the objective is convex, it is not quadratic.
The algorithm employed here solves its Lagrangian using an accelerated first-order method,
using only the objective and its gradient,
and is explained in detail in \cite{Jensen2012}.  An important technical detail for this algorithm
is that it requires that the image TV-term be differentiable. For the algorithm implementation
we use a smoothed TV-term, $\sum_j \sqrt{\| D_j \vec{f} \|_2^2 + \eta}$, with a small
smoothing parameter,
$\eta=10^{-10}$.
One convergence check on the algorithm is performed by evaluating
\begin{equation}
\label{cosa}
\cos \alpha =
 \frac{ \left(\nabla_{\vec{f}} R(\vec{f}) \right)
 \cdot \left( \nabla_{\vec{f}} D(X\vec{f}, \vec{g})  \right)}
{| \nabla_{\vec{f}} R(\vec{f}) |  | \nabla_{\vec{f}} D(X\vec{f}, \vec{g})|},
\end{equation}
where $R(\vec{f})$ denotes a generic regularization term,
and for constrained TV-minimization $R(\vec{f})=\| \vec{f} \|_{TV}$.
The conditions for convergence, derived in Ref. \cite{sidky2008image},
are that the gradients of the data-error and regularization
terms are back-to-back, $\cos \alpha = -1$, and $D(X\vec f, \vec g)=\epsilon$.
The latter condition assumes that the data-error constraint is active, which is the
case for all the simulations performed here.
For the present results, iteration is terminated when both
\begin{align}
\label{convcond}
\cos \alpha & \le -0.9999 \\
|D(X\vec{f}, \vec{g}) - \epsilon|/\epsilon & \le 0.001 \notag
\end{align}
are satisfied.

\subsection{Admitted undersampling by \ltwotv{}}
\label{sec:ltwoundersampling}

\begin{figure*}[!t]
\begin{minipage}[b]{0.5\linewidth}
\centering
\centerline{\includegraphics[width=\linewidth]{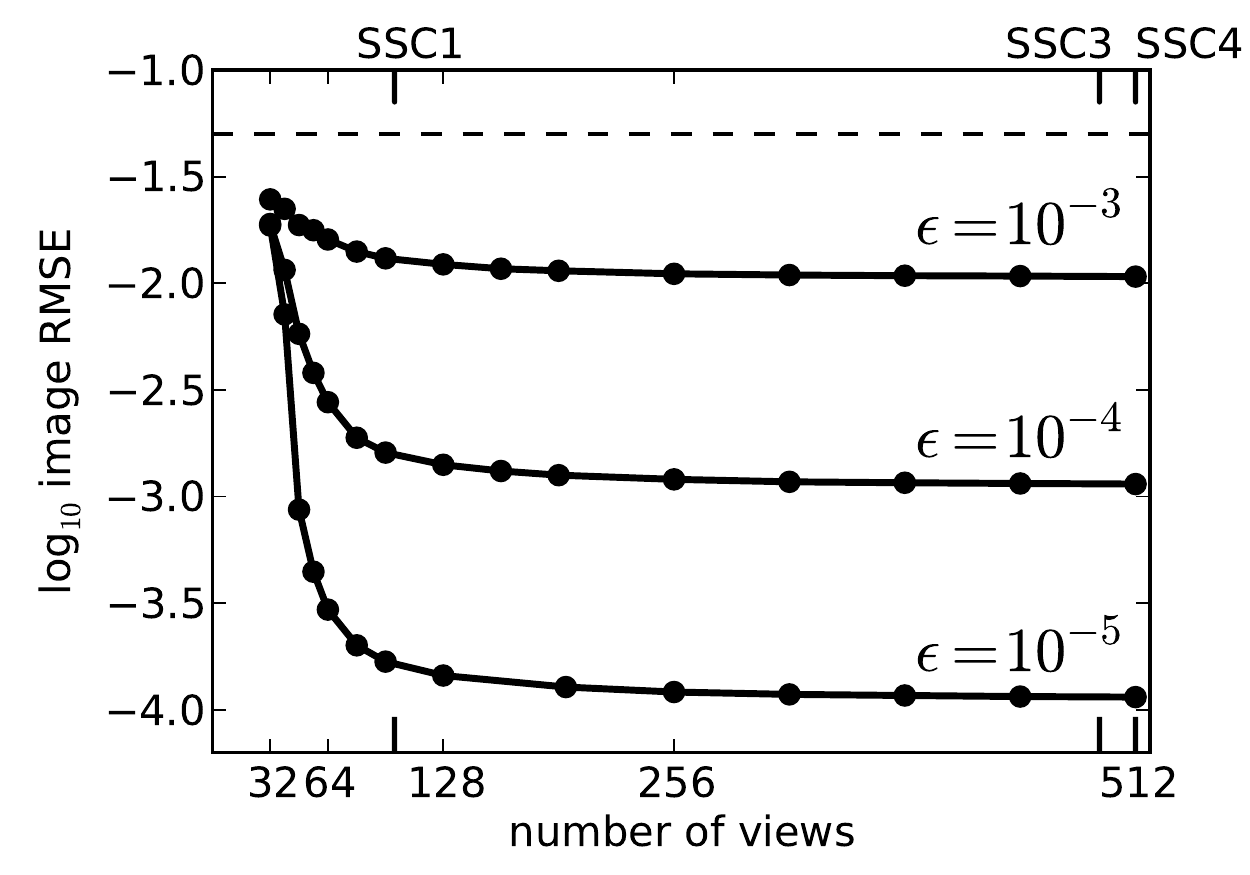}}
\end{minipage}
\begin{minipage}[b]{0.5\linewidth}
\centering
\centerline{\includegraphics[width=\linewidth]{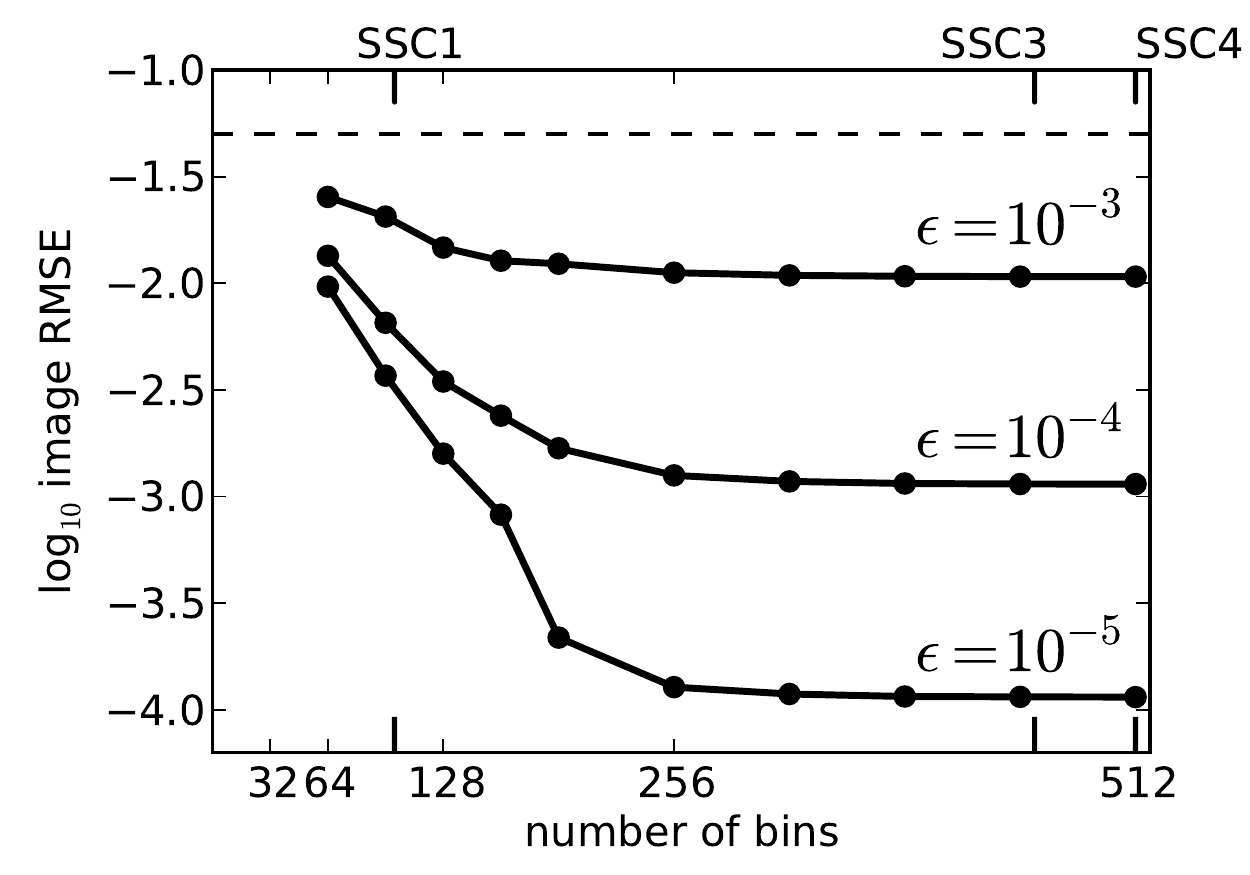}}
\end{minipage}
\caption{Image RMSE, $\delta$, for the data distance $D(X\vec f, \vec g)$ is constrained by $\epsilon = 10^{-3}, 10^{-4}, \text{ and } 10^{-5}$. The image size is $256 \times 256$.
Left: fixed $\nb = 2N = 512$, as function of $\nv$. Right: fixed $\nv = 2N = 512$, as function of $\nb$. 
The labels SSC1, SSC3 and SSC4 are the sufficient-sampling conditions discussed in Sec. \ref{sec:ssc_def}.
}
\label{fig:differentepsilons}
\end{figure*}

We are interested in two separate notions of accurate reconstruction: \emph{exact reconstruction} and \emph{stable reconstruction}. By the former, we mean that the reconstructed image is identical to the original. Exact reconstruction is only possible when $\epsilon = 0$, because the regularizing effect of a non-zero $\epsilon$ introduces a bias relative to the original image. Having $\epsilon = 0$ is only relevant for noise-free data, which means that stability is not an issue. Since SSC2 ensures a unique solution to \eqref{xfeg}, it can be used as a full sampling reference point for exact reconstruction. In practice, for the considered system matrix class, we found little difference in locations of SSC1 and SSC2, and we will therefore use SSC1 as a surrogate for SSC2.

Stable reconstruction is the corresponding concept for fixed, non-zero $\epsilon$, where we cannot hope for exact reconstruction. Instead, we are interested in the degree of sampling at which further increase in sampling leads to no further improvement in the reconstruction. This point of stable reconstruction can be compared to SSC3, since that is the point where no further improvement of the system matrix condition number occurs.
Since SSC4 was seen to approximate SSC3 for the present system matrix class and is simple to determine, it could also be considered to use SSC4 instead for the reference point.

In the simulations, we take the image array
to be the same as that of the breast phantom $N=256$.
The parameters of the circular, fan-beam
system matrix class are varied in a fashion parallel to the condition number
plots of Fig. \ref{fig:cond_views_bins}: first, $\nb$ is fixed at $2N$ and
$\nv$ is varied in the range $[32,512]$; and second, $\nv$ is fixed at $2N$ and 
$\nb$ is varied in the range $[32,512]$.

Computing SSC1 and SSC4 is straightforward and they occur at
\hbox{$\nv = 101$} and $\nv = 512$, respectively. 
Computation of SSC3 was performed
using the procedure outlined in Sec. \ref{sec:ssc_larger}. For the fixed-bin case, SSC3 occurs at $\nv = 492$, with $r_\text{samp} = 1.51$ and for the fixed-view case at $\nb = 456$.

In order to assess the undersampling with respect to exact reconstruction admitted by exploiting image gradient-magnitude
sparsity, we need access to the solution of \ltwotv{} for a data-equality constraint, $\epsilon=0$.
To our knowledge, solving \ltwotv{} for $\epsilon=0$ accurately with the present system
size is currently impractical. Instead we solve \ltwotv{} for $\epsilon = 10^{-3}, 10^{-4}, \text{ and } 10^{-5}$
to study the reconstruction error as $\epsilon$ approaches zero.
As an error measure, we use the root-mean-square-error (RMSE),
\begin{equation}
\label{iRMSE}
\delta = \|\vec{f}^* - \vec{f}_0\|_2 / \sqrt{N_\text{pix}},
\end{equation}
where $\vec{f}^*$ is the solution to \ltwotv{} and 
$\vec{f}_0$ is the original phantom.
%

The computed RMSEs for the results from \ltwotv{}
are displayed in Fig. \ref{fig:differentepsilons}.
%
As in Sec. \ref{sec:svd} we show SSC-locations at the top and bottom.
The horizontal line shows the minimum gray
level contrast, $0.05$, in the test phantom and provides a reference for the RMSE. An image
RMSE much less than the minimum gray-level contrast is an indicator that the reconstructed
image is visually close to the original phantom (barring pathological distributions of the image error).

For the plots of $\delta$ versus $\nv$,
we note a steep drop in  $\delta$ as $\nv$ increases past 40 views and the drop is increasingly
rapid as $\epsilon$ decreases. Based on these curves 
we extrapolate that exact reconstruction would be attained for $\nv \approx 50$
at $\epsilon=0$. Because SSC1 occurs at $\nv=101$, we note an admitted undersampling with
respect to 
exact reconstruction 
of a factor
of $2$ for the present simulation.
Note that use of SSC1 leads to a conservative estimate, because
SSC2 can only be larger than SSC1.

\begin{figure}[!t]
\includegraphics[width=\linewidth]{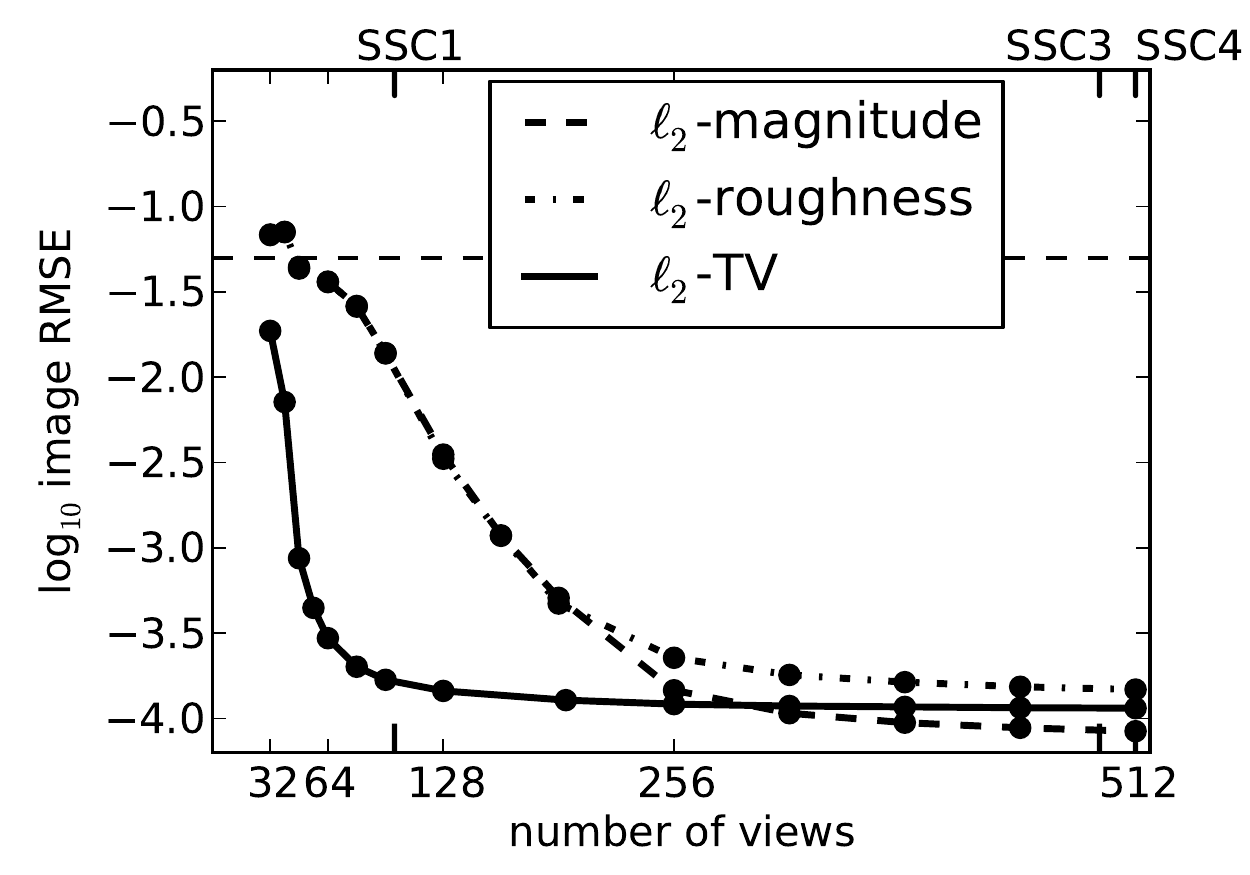}
\caption{
Image RMSE, $\delta$, for \ltwomag{}, \ltworough{} and \ltwotv{} reconstructions 
as a function of the number of views 
for the number of bins fixed at
$512$, and the data distance $D(X\vec f, \vec g)$ constrained by 
$\epsilon = 10^{-5}$. 
The horizontal dashed line shows the level of the minimum gray-level contrast
in the breast phantom. The labels
SSC1, SSC3 and SSC4 are the sufficient-sampling conditions discussed in Sec. \ref{sec:ssc_def}.
\label{fig:CSimageError}}
\end{figure}

For the plots of $\delta$ versus $\nb$, the image RMSE curves drop much more gradually
at each of the $\epsilon$'s investigated. Based on these curves it is only clear that $\delta$
is tending to zero at $\nb=190$ as function of $\epsilon$.  
Comparing to SSC1 at $\nb=101$, 
we do not observe any level of admitted undersampling in the bin-direction with respect to 
exact reconstruction.
Extending the range of $\epsilon$ to smaller values may yield a different conclusion.

This difference reflects an
asymmetry in sampling of the two parameters of $X$. We note that the asymmetry
was also observed in the condition number dependence on $\nv$ and $\nb$ for
the $N=32$ simulations in Sec. \ref{sec:smallsys}. For the present $N=256$ simulations, a relatively large
$\kappa(X) = 3.2\cdot 10^4$ 
is seen for $\nb=128$ and  $\nv=512$, compared to 
$\kappa(X) =  1.5\cdot 10^3$
for $\nb=512$ and $\nv=128$.
The results demonstrate a larger potential for successful
TV-based reconstruction from few views compared to using few bins.


Regarding stable reconstruction, 
%
we note that the
curves in Fig. \ref{fig:differentepsilons} all exhibit a plateau, where $\delta$
levels off with increasing $\nv$ or $\nb$, meaning that no gain in image RMSE is achieved by
increasing sampling. Thus, the left-most point of these plateaus is the point of stable reconstruction.
For the plot varying $\nv$, we see that 
stable reconstruction 
begins at $\nv \approx 80$,
which is a factor of $6$
fewer than SSC3.
For the plot varying $\nb$, the 
stable reconstruction 
begins at $\nb \approx 200$, 
a factor of approximately $2$ fewer than SSC3.

These results show \emph{quantitatively} that significant undersampling in $\nv$, particularly with respect to stable reconstruction,
is admitted for \ltwotv{}.  
This conclusion is achieved with a phantom
modeling a realistic level of gradient-magnitude sparsity. We do point out, however, that these empirical
results only apply to the presented simulation.
To support the present conclusion for admitted undersampling, we vary in the following sections different aspects of the \ltwotv{} study.

\subsection{Altering the optimization problem}

\begin{figure*}[!t]
\begin{minipage}[b]{0.5\linewidth}
\centering
\centerline{\includegraphics[width=\linewidth]{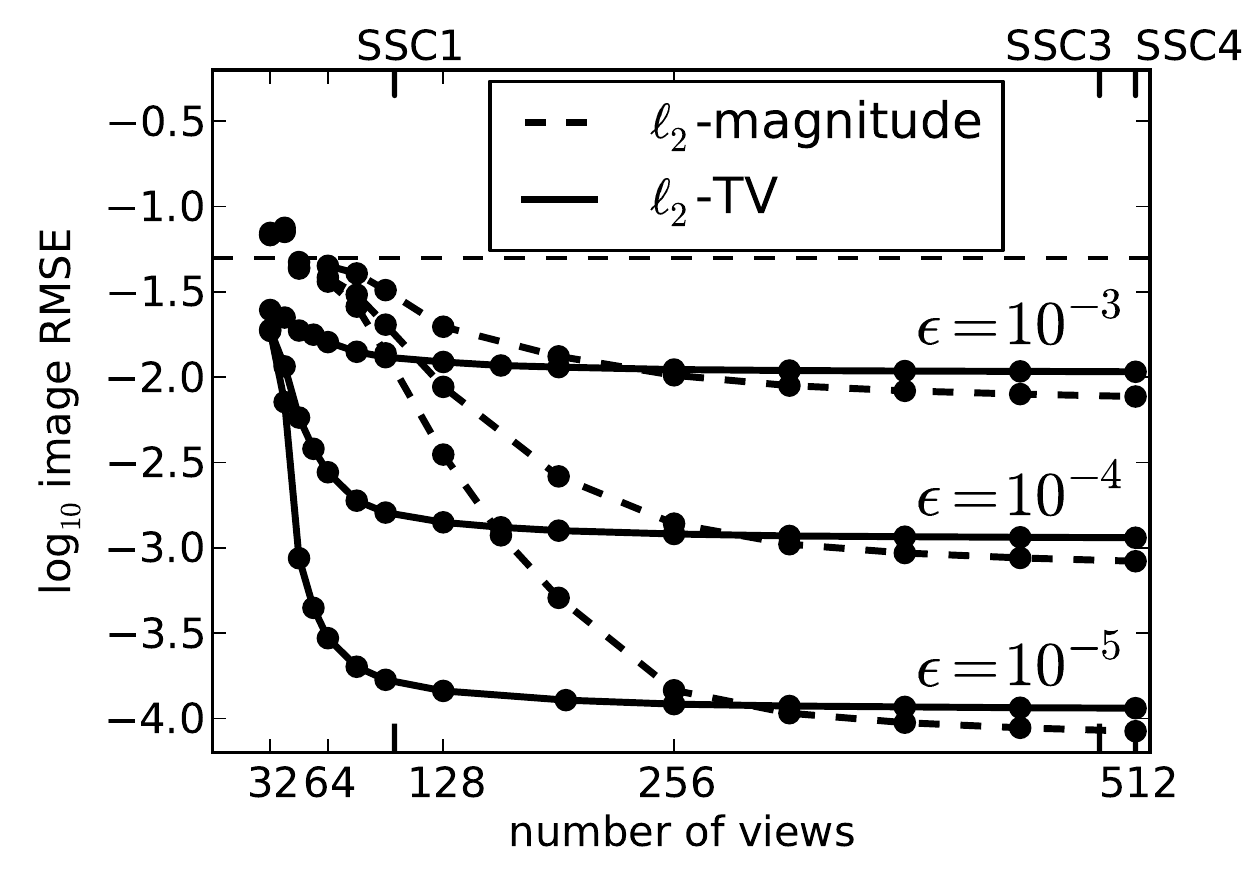}}
\end{minipage}
\begin{minipage}[b]{0.5\linewidth}
\centering
\centerline{\includegraphics[width=\linewidth]{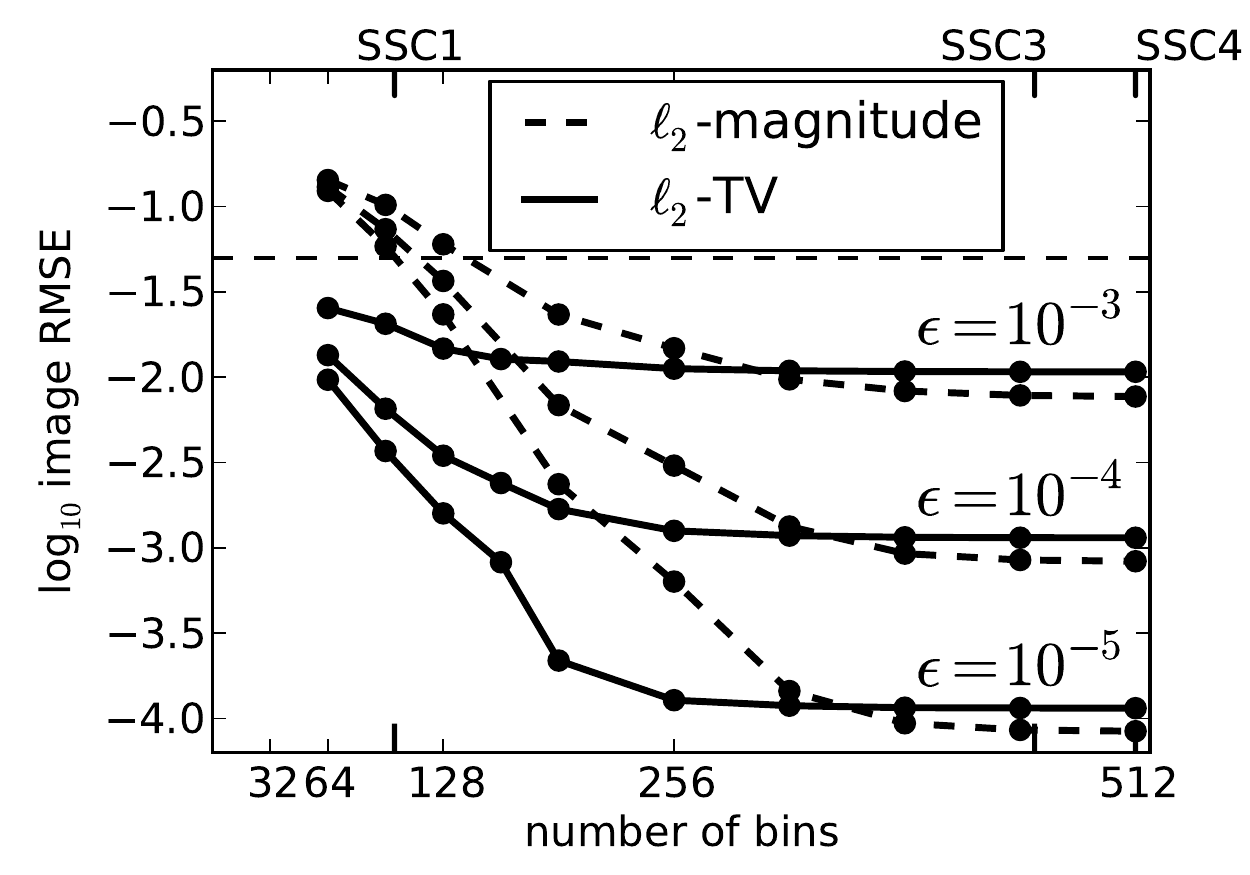}}
\end{minipage}
\caption{Image RMSE, $\delta$, for \ltwomag{} and \ltwotv{} reconstructions
and $\epsilon = 10^{-3}, 10^{-4}, \text{ and } 10^{-5}$.
Left: fixed $\nb = 2N = 512$, as function of $\nv$. Right: fixed $\nv = 2N = 512$, as function of $\nb$.
The horizontal dashed line shows the level of the minimum gray-level contrast
in the breast phantom. 
The labels SSC1, SSC3 and SSC4 are the sufficient-sampling conditions discussed in Sec. \ref{sec:ssc_def}.
\label{fig:differentepsilonsAll}}
\end{figure*}

To support the use of the gradient-magnitude sparsity exploiting \ltwotv{}
for admitting undersampling, we compare results
with two other optimization problems,
\begin{align}
\,\text{\ltwomag{}:} \qquad &\vec{f}^* = \argmin_{\vec{f}} \| \vec{f} \|_2^2  \label{eq:ltwomag} \\ 
\text{subject to} \qquad &D(X\vec f, \vec g) \leq \epsilon, \qquad\quad \notag
\end{align}
and
\begin{align}
\text{\ltworough{}:} \qquad &\vec{f}^* = \argmin_{\vec{f}} \| \nabla \vec{f} \|_2^2 \label{eq:ltworough} \\
\text{subject to} \qquad &D(X\vec f, \vec g) \leq \epsilon, \qquad\;\; \notag
\end{align}
where $\nabla$ represents a numerical gradient operation and is computed by forward finite-differencing.
The Lagrangian form of these optimizations are two forms of Tikhonov regularization commonly used for IIR.

The solutions to \ltwomag{} and \ltworough{} are obtained with linear conjugate gradients
(CG) applied to the Lagrangians of these problems with the multiplier being adjusted
until the data-error constraint holds with equality. The convergence criteria are the same
as what is specified in Eq. (\ref{convcond}) except that $R(\vec{f}) = \|\vec{f}\|^2_2$
for \ltwomag{} and $R(\vec{f}) = \| \nabla \vec{f}\|^2_2$
for \ltworough{}.

We focus on the $\epsilon = 10^{-5}$ case and plot image RMSEs for \ltwomag{}, \ltworough{} and \ltwotv{} as function of $\nv$ for fixed $\nb = 512$ in Fig. \ref{fig:CSimageError}.
We note little difference between results from \ltwomag{} and \ltworough{} but a large gap between these results
and those of \ltwotv{}. 
The optimization problems \ltwotv{} and \ltworough{} differ
only on the norm of the image-gradient in the regularization term, while \ltworough{} and \ltwomag{} differ by the presence
of the gradient. It is clear from Fig. \ref{fig:CSimageError} that for this simulation, the $\ell_1$-norm has the greater
impact.

For large $\nv$,
the \ltwotv{} RMSE is actually slightly inferior to that of \ltwomag{}.
The reason is the regularizing effect of having a non-zero $\epsilon$, which causes a
small bias of the solutions compared to the original image. The relative size of the biases are not
known in advance.
We conclude that the \ltwotv{} solution is not 
to prefer 
over the \ltwomag{} 
and \ltworough{} solutions when $\nv$ approaches the SSC3 ($r_\text{samp} = 1.5$).
Nevertheless, there is a certain ``sampling window'',
for the present phantom, approximately for $\nv \in [50, 256]$,
where the constrained TV-minimum solution is superior to Tikhonov regularization in terms of RMSE. 

In Fig. \ref{fig:differentepsilonsAll}, we overlay the results of \ltwomag{} onto the results
of \ltwotv{} from Fig. \ref{fig:differentepsilons} in order to investigate possible undersampling
admitted by \ltwomag{}. The results of \ltworough{} are not shown because they are similar
to those of \ltwomag{} and to prevent clutter in the figure. Going from left to right,
both plots show a gradual decrease
of $\delta$ for \ltwomag{} as $\nv$ and $\nb$ increase with $\delta$ leveling off at $\nv \approx300$
and $\nb \approx 400$. 
For the investigated range of $\epsilon$, \ltwomag{} does not admit any undersampling 
with respect to exact reconstruction, but does show a marginal undersampling with respect to stable reconstruction as the corresponding
$\delta$-curves reach the plateau before SSC3 and SSC4. 
In summary, 
the undersampling admitted by \ltwomag{} is substantially
less that that admitted by \ltwotv{} for this simulation; particularly in considering
view number undersampling with respect to stability.

\subsection{Altering the image evaluation metric}
\label{sec:images}

\begin{figure}[!t]
\includegraphics[width=\linewidth]{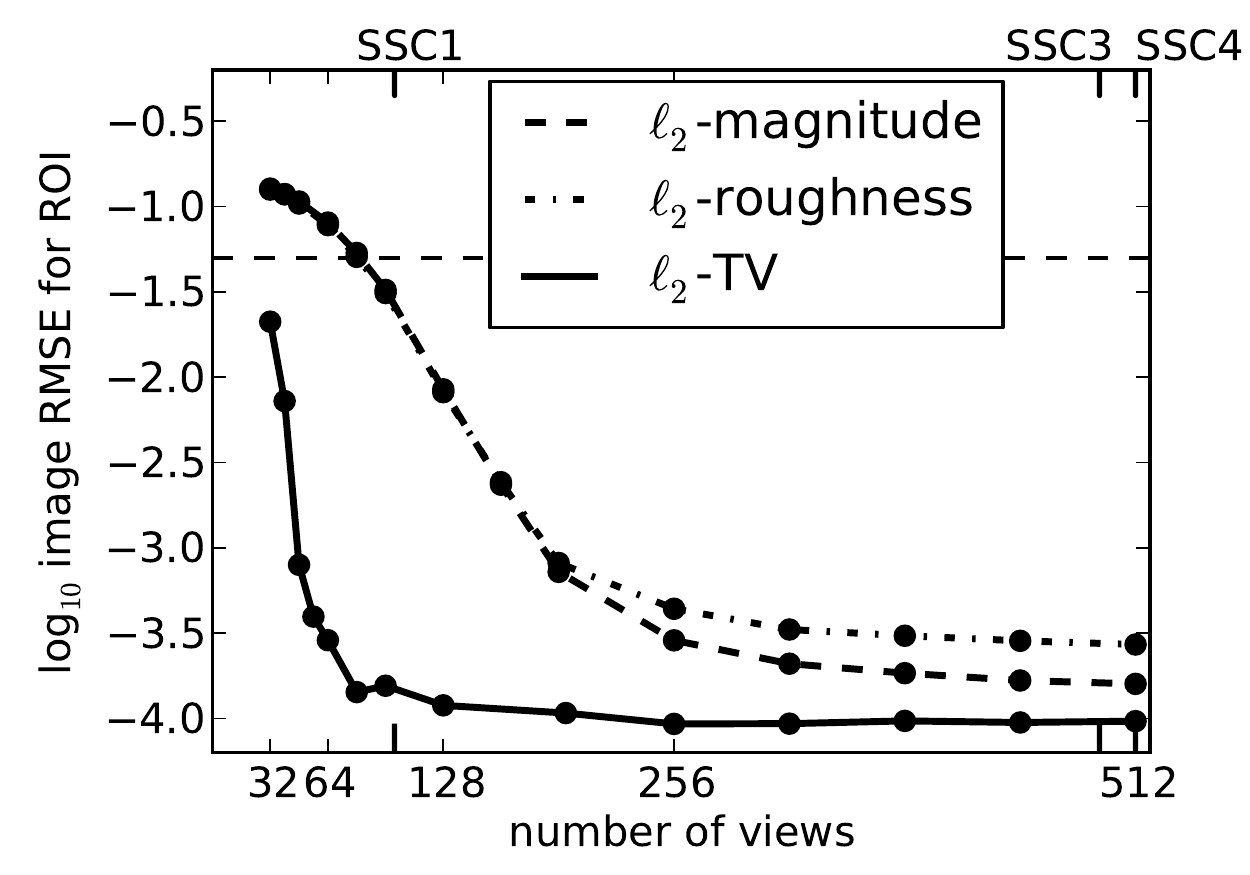}
\caption{
Image RMSE as in Fig. \ref{fig:CSimageError} but for $\delta_\text{ROI}$, i.e., the RMSE restricted to the ROI around the microcalcifications.
\label{fig:CSimageErrorROI}}
\end{figure}

Conclusions based on evaluating reconstructed images with a single summarizing metric, such as the RMSE, can be misleading.
While our aim is not a fully realistic image evaluation, we want to show how the results can potentially change with a change of metric. 
For example, with the task of microcalcification detection in mind, one might consider the RMSE of only the ROI of the microcalcifications displayed in Fig. \ref{fig:breastPhantom}. This RMSE is denoted $\delta_\text{ROI}$.
We show the corresponding plot
for $\delta_\text{ROI}$ in Fig. \ref{fig:CSimageErrorROI}. While there are numerical differences
between the $\delta$ and $\delta_\text{ROI}$ plots, the trends are similar giving us further confidence
that the RMSE of the entire image, $\delta$, can be used for investigating admitted undersampling.

%
\begin{figure*}[!t]
\begin{minipage}[b]{\linewidth}
\centering
\centerline{\includegraphics[width=\linewidth]{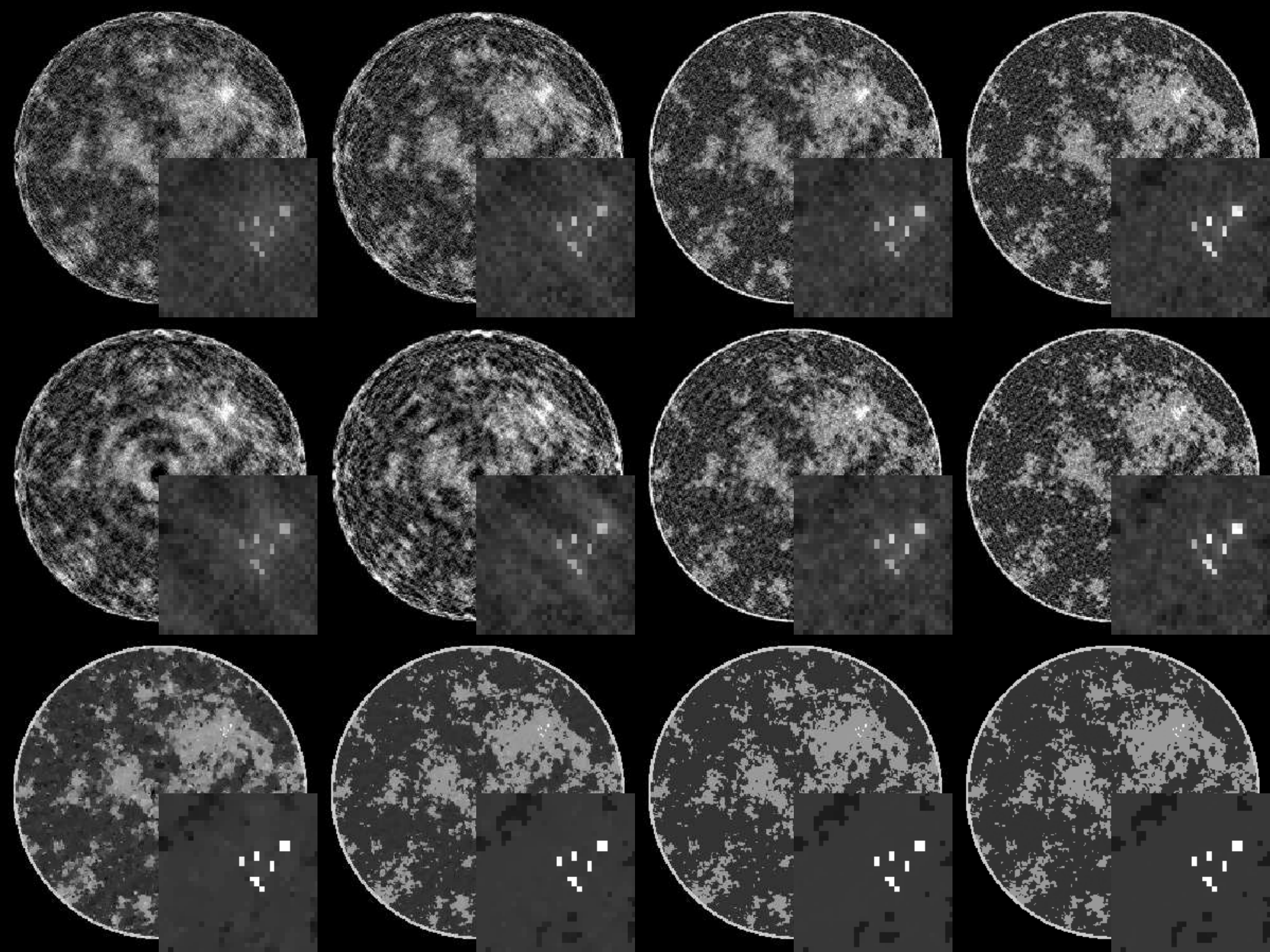}}
\centerline{\includegraphics[width=\linewidth]{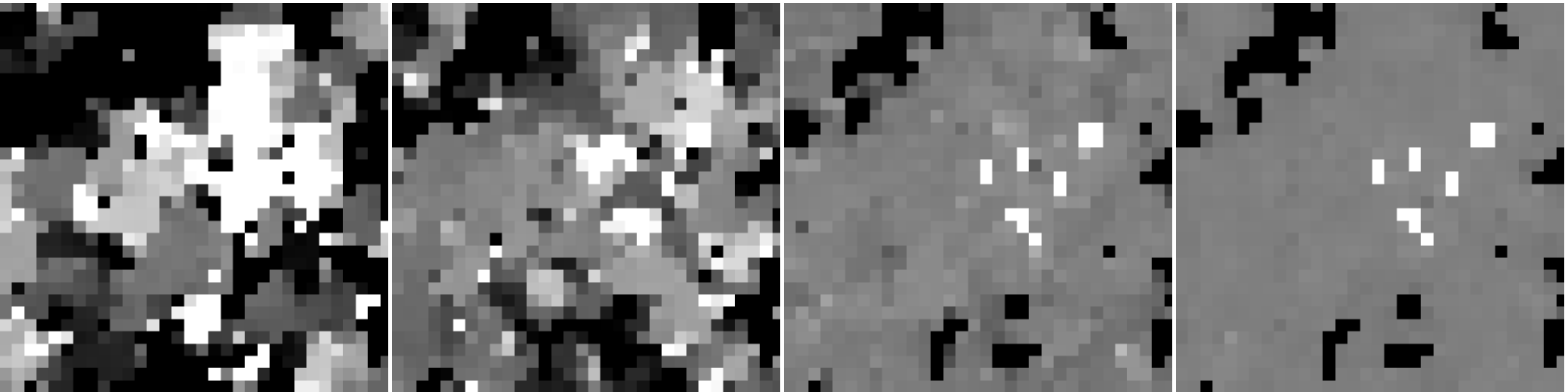}}
\end{minipage}
\caption{1st, 2nd, 3rd and 4th columns
show reconstructions from $32$, $40$, $48$, $64$ view data. The 1st, 2nd and 3rd rows
show \ltwomag{}, \ltworough{} and \ltwotv{} images with $\epsilon = 10^{-5}$.
The gray scale window is $[0.95, 1.20]$ for the complete image, and $[0.9, 1.8]$ for the ROI blow-ups.
The 4th row shows the \ltwotv{} ROIs enlarged in an extremely narrow gray scale, $[1.09, 1.11]$,
in order to scrutinize the transition to sufficient sampling based on the
object sparsity. These images show that $50$ views is sufficient
for this system and object.
\label{fig:CSimages}}
\end{figure*}

Another way to evaluate images is by visual comparison.
The reconstructed images in Fig. \ref{fig:CSimages} are shown for a range in $\nv$ showing
the transition to accurate image reconstruction by \ltwotv{}. That the \ltwomag{} and \ltworough{}
show strong artifacts for this range is expected as their corresponding image RMSEs are at
the level of the minimum phantom contrast level.
Interestingly, the microcalcifications can be identified and well-characterized in all reconstructions,
although more clearly with more views. It may be argued that, from a utility point of view,
that $32$ views would suffice if we are solely
interested in the microcalcifications and disregard the prominent artifacts of the background image. 
The ROI of the \ltwotv{} reconstructed images are
shown with a narrow gray scale window in the bottom row to reveal the high level of accuracy at $\nv=50$.
We emphasize here that our goal is not to go into a discussion about different artifacts but simply
support our conclusions on undersampling from Sec. \ref{sec:ltwoundersampling} by illustrating the behavior in
the transition region around $\nv = 50$.

\subsection{Altering the system matrix class}

\begin{figure}[!t]
\begin{minipage}[b]{\linewidth}
\includegraphics[width=\linewidth]{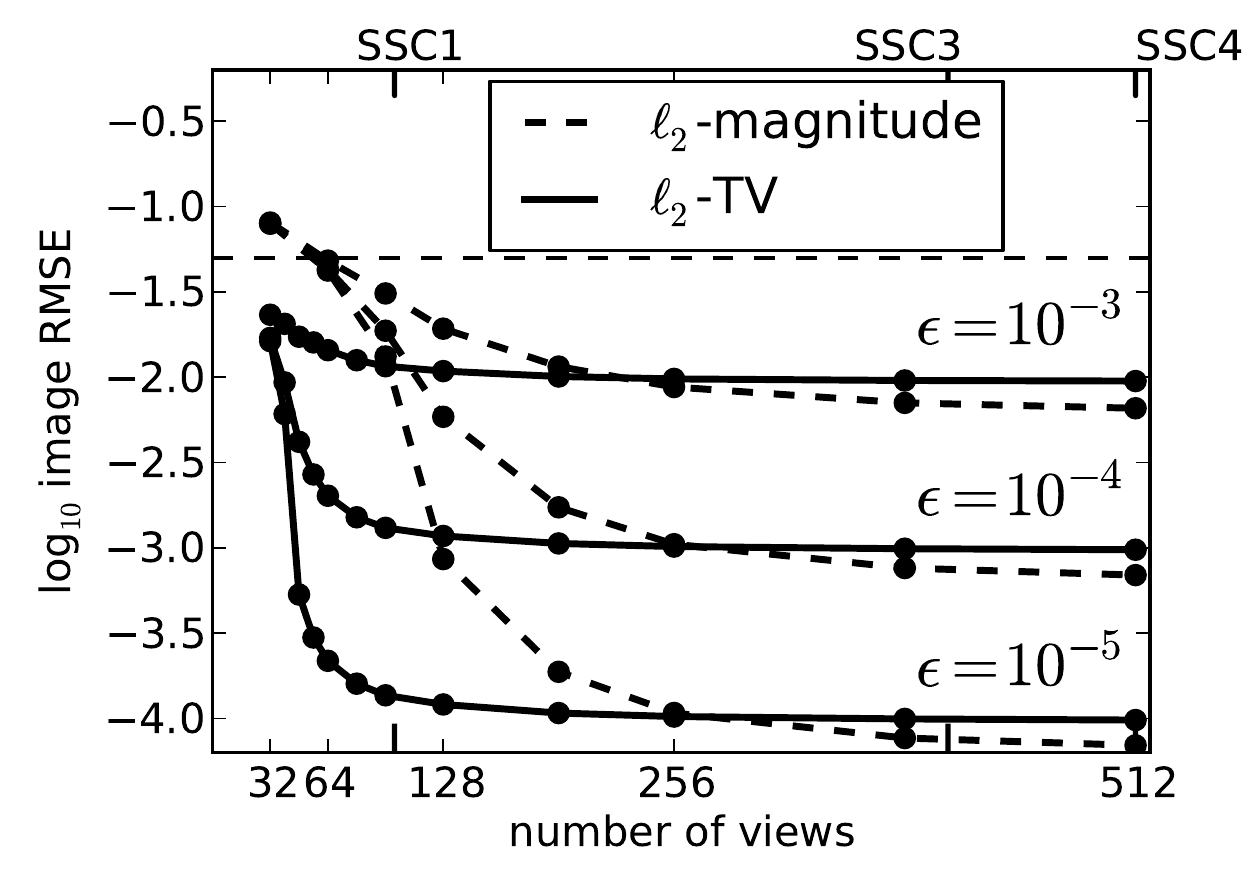}
\end{minipage}
\caption{
Image RMSE, $\delta$, for \ltwomag{} and \ltwotv{} reconstructions,
as in Fig. \ref{fig:CSimageError} for varying $\epsilon$, computing $X$ with
ray-tracing and nearest-neighbor interpolation. $\nb$ is fixed at $512$. The labels
SSC1, SSC3 and SSC4 are the sufficient-sampling conditions discussed in Sec. \ref{sec:ssc_def}.
}
\label{fig:differentepsilons_nn}
\end{figure}

To illustrate the change in results due to change in the 
system matrix 
class,
the RMSE, $\delta$, is again computed as function of $\nv$ for $\nb = 2N = 512$ and
for $\epsilon = 10^{-3}$, $10^{-4}$ and $10^{-5}$ but using a system matrix set
up through ray-tracing with nearest-neighbor interpolation at the mid-line
of each pixel row, as in Sec. \ref{sec:ssc_alternate_model}. Results are
plotted in Fig. \ref{fig:differentepsilons_nn}. The overall trends are
similar to those for line-intersection, however SSC3 occurs already at $\nv = 408$,
with $r_\text{samp} = 1.49$. By closer comparison with the line-intersection
results, it is seen that the nearest-neighbor RMSEs are smaller
than the line-intersection RMSE at the same $\nv$.
This example serves to illustrate that the 
SSCs will change when the 
system matrix 
class is altered.  

While the present
alteration is relatively minor, it is enough that the approximation SSC4 of to SSC3 is worse, and larger differences can be expected with
more radical changes such as the use of non-point-like image expansion functions.

In terms of admissible undersampling for \ltwotv{} with the altered system matrix class, we see very similar undersampling factors for both exact and stable reconstruction as for the line-intersection class. 

\subsection{Altering the phantom sparsity}
\begin{figure*}[!t]
\begin{minipage}[b]{0.666\linewidth}
\centering
\centerline{\includegraphics[width=0.75\linewidth]{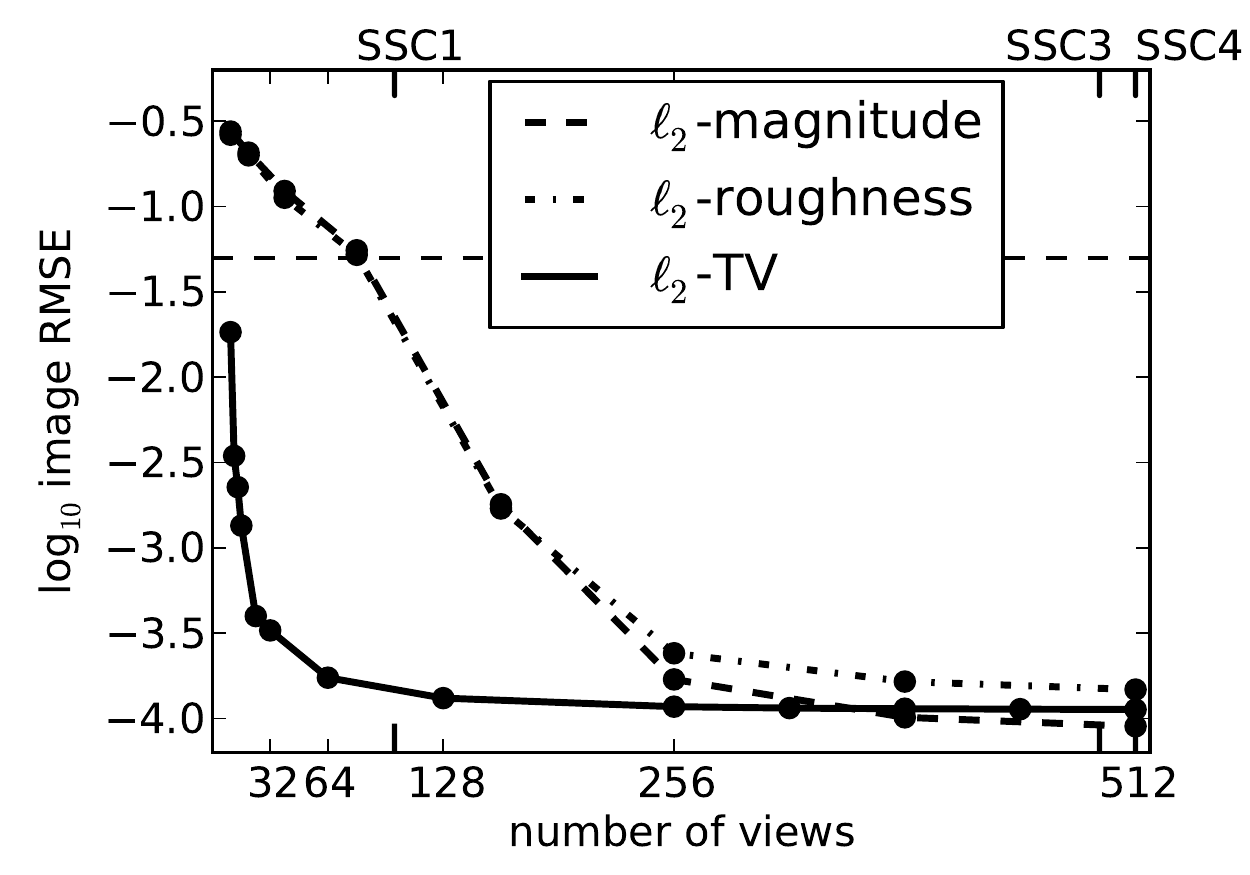}}
\end{minipage}
\begin{minipage}[b]{0.33\linewidth}
\centering
\centerline{\includegraphics[width=\linewidth]{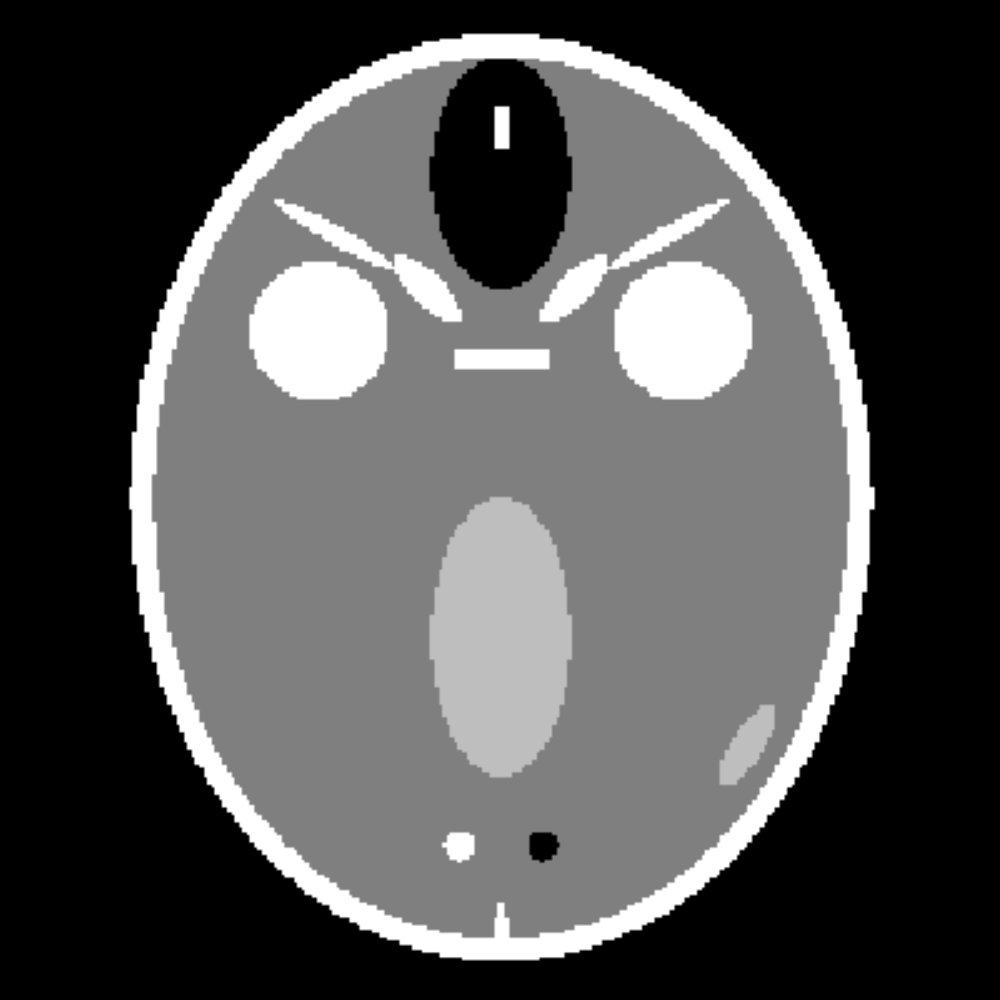}}
\vspace*{2mm}
\end{minipage}
\begin{minipage}[b]{\linewidth}
\centering
\centerline{\includegraphics[width=\linewidth]{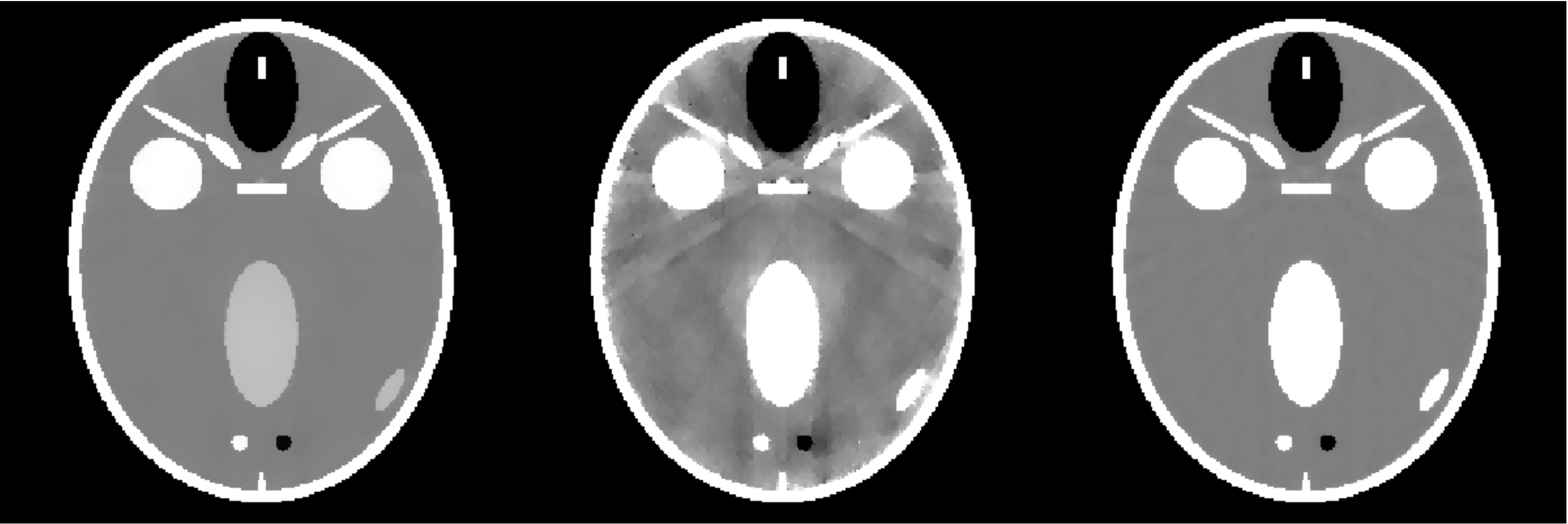}}
\end{minipage}
\caption{
Top: 
Image RMSE, $\delta$, for \ltwomag{}, \ltworough{} and \ltwotv{} reconstructions,
as in Fig. \ref{fig:CSimageError}, 
except that
the data are generated from the head phantom shown on the right. The labels
SSC1, SSC3 and SSC4 are the sufficient-sampling conditions discussed in Sec. \ref{sec:ssc_def}.
Bottom: images reconstructed for $N_\text{views} = 12$ shown in a gray
scale window of $[0.9,1.1]$ on the left and a narrow gray scale window of $[0.99, 1.01]$
in the middle. On the 
right is the reconstructed image for $\nv=32$ in the
narrow gray scale window $[0.99,1.01]$. At $N_\text{views} = 12$ the RMSE is $0.005$ resulting
in visible artifacts for the image shown in the $1 \%$ gray scale window, while the RMSE
is a factor of $10$ less for $\nv =32$.
\label{fig:head}}
\end{figure*}

The breast phantom study is repeated employing
a variation of the FORBILD head phantom \cite{forbildhead} 
which is highly sparse in the gradient-magnitude image.
The present version of the phantom, which is seen in Fig. \ref{fig:head}, does not
have the ear objects of the original phantom, and the contrast levels have been increased
so that the minimum gray-level contrast is the same as for the breast phantom.
The gradient magnitude
sparsity is $2,492$, or approximately a quarter of the breast phantom.
In Fig. \ref{fig:head}, the obtained $\delta$ for \ltwomag{},
\ltworough{} and \ltwotv{} are shown as function of $\nv$ with $\nb = 2N=512$ and $\epsilon = 10^{-5}$.

The \ltwomag{} and \ltworough{} curves are almost identical to those of the breast phantom. The SSCs
are 
unchanged, because the same 
system matrix class
is used. That two so different looking phantoms show such similar $\delta$-behavior
suggests that the reconstruction quality of \ltwomag{} and \ltworough{} 
depend only weakly on the particular phantom. 

For \ltwotv{}, on the other hand,
the step-like transition occurs already at $\nv = 12$, for which the reconstruction
is shown in Fig \ref{fig:head}. We expect that this phantom
would be recovered exactly at $\nv=12$ in the limit $\epsilon \rightarrow 0$, leading to an admitted undersampling with respect to exact reconstruction of a factor of approximately $8$. 
Stable reconstruction occurs at $\nv\approx 64$, i.e., an undersampling also of a factor $8$ with respect to stability.

Interestingly, the exact reconstruction result hints at the existence of a simple relation between sparsity and admitted undersampling. 
Compared to the breast phantom, there is a gain in undersampling by $8 / 2 = 4$. In comparison, we note that the change in gradient magnitude sparsity relative to the breast phantom is $10,019 / 2,492 \approx 4$. That is, reducing the sparsity by a certain factor leads to an improvement in the admitted undersampling by the same factor.
%
%
This result, if shown to hold, could be important for practical
application of CS-inspired sparsity-exploiting methods, since it provides quantitative insight into how many views would suffice for reconstructing images of given sparsity. 
%
Another conclusion that can be drawn is that simulations with images of too low sparsity compared to a realistic level in the imaging scenario of interest are bound to yield over-optimistic promises of undersampling potential. 
This could have been anticipated but the result establishes this expectation quantitatively.
%

We caution, however, that the
result is based on only two phantoms and further study is required.
For more robust conclusions, the present studies need to be performed on ensembles of
phantoms in order to verify that admitted undersampling 
for constrained 
TV-minimization depends primarily on the gradient-magnitude image sparsity.
We also note that while we may have exact reconstruction of the head phantom at $\nv = 12$ and the reconstructed image at $\epsilon = 10^{-5}$ appears very accurate in the $[0.9, 1.1]$ gray scale window, it is in fact \emph{not} an exact reconstruction. By narrowing the gray scale to $[0.99, 1.01]$, also shown in Fig. \ref{fig:head}, prominent artifacts become visible. This underlines that exact reconstruction is not the relevant notion for a fixed, non-zero $\epsilon$. Instead, stable reconstruction, at $\nv=64$, yields an accurate reconstruction, and for the present case already at $\nv = 32$ (also shown in Fig. \ref{fig:head}) the artifacts are reduced to a negligible level in the $[0.99,1.01]$ gray scale window.

\section{Summary}
We argue that
a quantitative notion of a sufficient-sampling condition
(SSC) for X-ray CT using the DD imaging model is necessary in order to provide
a reference for evaluating the undersampling
potential of sparsity-exploiting methods.
We propose and apply four different SSCs to a 
class of system matrices describing
circular, fan-beam CT with a pixel expansion. 
While SSC1 and SSC2 
characterize invertibility of the system matrix, 
SSC3 characterizes numerical stability for inversion of the system matrix.
A simple-to-compute SSC4 is seen to approximate SSC3 closely for the circular, fan-beam full angular range CT geometry.

We employ the SSCs as reference points of full sampling to quantify undersampling admitted by reconstruction through TV-minimization on a breast CT simulation.
Relative to SSC1, we observe some undersampling potential of TV-minimization for exact reconstruction and large undersampling relative to SSC3 and SSC4 for stable reconstruction. We find few-view reconstruction to admit larger undersampling than few-detector bin reconstruction, and we show evidence of a simple quantitative relation between image sparsity and the admitted undersampling.


More generally, the proposed SSCs can help to engineer and understand other sparsity-exploiting IIR algorithms 
by providing full sampling reference points for the system matrix class associated with the imaging application of interest in order to quantify the admissible undersampling.
This analysis can guide decisions on alternative optimization problems, object representations, sampling configurations,
and integration models.


\section*{Acknowledgment}
This work is part of the project CSI: Computational Science
in Imaging, supported by grant 274-07-0065 from the Danish
Research Council for Technology and Production Sciences.
This work was supported in part by NIH R01 grants CA158446, CA120540, and EB000225.
The contents of this article are
solely the responsibility of the authors and do not necessarily
represent the official views of the National Institutes of Health.

\ifCLASSOPTIONcaptionsoff
  \newpage
\fi



\bibliographystyle{IEEEtran}
\bibliography{sampling}

%








\end{document}